\documentclass[a4paper,11pt]{article}
\usepackage{jheppub} 
\usepackage{lineno}
\usepackage{enumitem}
\usepackage{cleveref}
\usepackage{amsmath}
\usepackage{dsfont}
\usepackage{placeins}
\usepackage{braket}
\usepackage{subcaption}
\usepackage[dvipsnames]{xcolor}
\newcommand{\ph}{\varphi}

\title{ Probing the self-coherence of primordial quantum fluctuations with complexity}

\author[a]{Arpan Bhattacharyya,}
\author[b]{Suddhasattwa Brahma,}
\author[c,d]{S. Shajidul Haque,}
\author[c]{Jacob S. Lund,}
\author[e]{and Arpon Paul}
\affiliation[a]{Indian Institute of Technology, Gandhinagar, \\Gujarat-382355, India}
\affiliation[b]{Higgs Centre for Theoretical Physics, School of Physics and Astronomy, University of Edinburgh,
Edinburgh EH9 3FD, UK}
\affiliation[c]{Department of Mathematics and Applied Mathematics, University of Cape Town,\\ Cape Town-7700, South Africa}
\affiliation[d]{National Institute for Theoretical and Computational Sciences (NITheCS), Private Bag X1, Matieland,
South Africa}
\affiliation[e]{School of Physics and Astronomy, University of Minnesota,\\ Minneapolis, Minnesota 55455, USA}

\emailAdd{abhattacharyya@iitgn.ac.in}
\emailAdd{suddhasattwa.brahma@gmail.com}
\emailAdd{shajid.haque@uct.ac.za}
\emailAdd{jacob.lund1@gmail.com}
\emailAdd{paul1228@umn.edu}
\abstract{A smoking gun for our current paradigm of the early universe would be direct evidence for the quantum mechanical origin of density perturbations which are conjectured to seed the large scale structure of our universe. A recently-proposed novel phenomenon is that of \textit{recoherence}, wherein a specific interaction between the adiabatic and the entropic sector leads to the adiabatic mode retaining a coherent state after a transient increase in linear entropy. In this paper, we choose the Gaussian action allowing for both field-field and field-momentum coupling, and analyze the evolution of linear entropy, complexity of purification (COP), and complexity of formation (COF) to capture the interplay between decoherence and recoherence in this model. In the presence of these two types of couplings that drive these two opposing characteristics, we highlight how COF is an efficient tool for diagnosing dynamics for such an open quantum system.}

\begin{document}

\maketitle

\flushbottom
\section{Introduction}
In recent years, computational tools and methodologies developed for quantum information theory (QIT) have found transformative applications in fields such as quantum field theory, gravity, and cosmology. These approaches not only provide fresh perspectives but also inspire new questions that fundamentally reshape our understanding of complex quantum systems. Building on previous work, we will further utilize this framework to investigate the evolution of a cosmological model that will expand our understanding of the early-universe.

The origin of cosmological structures is commonly attributed to quantum fluctuations in the primordial vacuum, a mechanism central to both inflationary theory and many of its alternatives, \textit{e.g.,} \textit{Ekpyrosis} \cite{Lehners:2008vx}. However, we are yet to find direct signatures of the quantum nature of these fluctuations, which have clear consequences for data in the cosmic microwave background and the large scale structure of the universe. This is not only of purely academic interest -- the quantum origin of structure or the lack thereof will help establish the correct model of the early-universe\footnote{For instance, String Gas Cosmology \cite{brandenberger1989superstrings}, Matrix Cosmology \cite{Brahma:2021tkh}, and Warm Inflation \cite{Berera:1995ie} all invoke thermal fluctuations instead of quantum ones.}. However, the efficient nature of decoherence of the observable quantum fluctuations, due to the presence of environment degrees of freedom, typically erase genuine quantum signatures and our ability to find a smoking gun signal for them. Recent advancements in cosmological open quantum systems \cite{Brahma:2024yor, Brahma:2022yxu, Shandera:2017qkg} offer robust tools for exploring this phenomenon. Contrary to conventional expectations, \cite{Colas:2022kfu} demonstrate that the purity of the quantum state corresponding to adiabatic mode, when interacting with the entropic sector in a specific Gaussian mixing term, can have a transient decrease before bouncing back to culminate in a highly self-coherent final state. This phenomenon has no analogue in flat-space physics since it is due to the quenching of the interaction term, after a certain time, due to curved space effects. In this work, we investigate whether quantum complexity can serve as a tool to study this recoherence \cite{Colas:2022kfu, Burgess:2024eng}. This is a particularly relevant question in light of our recent work \cite{Bhattacharyya_2024}, where we have shown that certain features of quantum complexity can indeed signal decoherence -- or the loss of purity of the quantum state -- in an analogous cosmological system.

 The evolution of the Universe can be envisioned as a quantum circuit \cite{Bhattacharyya:2020rpy, Bhattacharyya:2020kgu, Haque:2021kdm, Haque:2021hyw, Lehners:2020pem, Liu:2021nzx, Saha:2022onq,Li:2023ekd}, where transitions between different phases of the evolution can be associated with the quantum complexity of each of these states. In this context, considering the primordial metric and matter fluctuations to be in a quantum state, the dynamics of complexity can provide valuable insights into the mechanisms of decoherence (and recoherence) of these degrees of freedom. One of the main motivations behind this lies in our lack of ability to track all the environment modes that cosmic observables interact with, and hence the requirement of considering the early universe as an open quantum system. Circuit complexity \cite{NL1,NL2,NL3}\footnote{Interested readers are referred to some of the works done in the context of Nielsen complexity \cite{NL1,NL2,NL3} for quantum field theory and quantum many-body systems \cite{Jefferson,Chapman:2017rqy,Bhattacharyya:2018wym,Caputa:2017yrh,Ali:2018fcz,Bhattacharyya:2018bbv,Hackl:2018ptj,Khan:2018rzm,Camargo:2018eof,Ali:2018aon,Caputa:2018kdj,Guo:2018kzl,Bhattacharyya:2019kvj,Flory:2020eot,Erdmenger:2020sup,Ali:2019zcj,Bhattacharyya:2019txx,Caceres:2019pgf,Bhattacharyya:2020art,Liu_2020,Susskind:2020gnl,Chen:2020nlj,Czech:2017ryf,Chapman:2018hou,Geng:2019yxo,Guo:2020dsi,Couch:2021wsm,Erdmenger:2021wzc,Chagnet:2021uvi,Koch:2021tvp,Banerjee:2022ime,Bhattacharyya:2022ren,Bhattacharyya:2023sjr,Bhattacharyya:2022rhm,Bhattacharyya:2021fii,Bhattacharyya:2020iic,Craps:2023rur,Jaiswal:2021tnt,Jaiswal:2020snm,Haque:2021kdm,Bhattacharya:2022wlp}. This list is by no means exhaustive and the reader is referred to the following reviews and thesis \cite{Chapman:2021jbh, Bhattacharyya:2021cwf,Katoch:2023etn,Aguilar-Gutierrez:2024yzu}, and references therein, for more details.} in open quantum systems has been extensively explored across various models \cite{Bhattacharyya:2021fii,Bhattacharyya_2024, Bhattacharyya:2022rhm} to understand its significance in quantifying the dynamics of the reduced density matrix corresponding to the system modes. Of particular interest is the study of decoherence in the flat-space Caldeira-Leggett model \cite{PhysRevLett.46.211}, as explored in \cite{Bhattacharyya:2022rhm}, where the time evolution of the complexity of purification (COP) \cite{Ali:2018fcz} was compared to other standard decoherence detection tools, such as linear entropy and entanglement negativity \footnote{For other measures of decoherence in the context of cosmology interested readers are referred to some of theses papers \cite{Martin:2021znx,Martin:2022kph}\,.}. Interestingly, this analysis could be successfully extended to curved spacetime, specifically for the two-field model in de Sitter background \cite{Bhattacharyya_2024}. For both flat and curved spacetime scenarios, circuit complexity demonstrated significant sensitivity to the decoherence mechanism. In light of this finding -- sensitivity of quantum complexity as a tool for decoherence -- it is natural to ask if complexity  can be  a quantifier for more exotic phenomena such as the recoherence mechanism established in \cite{Colas:2022kfu}.

An effective field theory approach to inflation treats terms coupling the hidden sector field to the shift-symmetric inflaton in a systematic way. At leading order, this leads to a mixing term between the heavy field being coupled to the momentum of the inflaton perturbation \cite{Assassi:2013gxa}. Starting with this Lagrangian, this model can be formulated in terms of a reduced density matrix formulation, where one field acts as the system and the other as the environment \cite{Colas:2022kfu, Colas:2024xjy, Brahma:2024ycc}. We adopt the same model to investigate whether quantum circuit complexity is sensitive enough to reflect this \textit{recoherent behavior}. Additionally, we include and examine field-field coupling to understand the distinctions in the characteristics of complexity between decoherence and recoherence. In this context, we demonstrate that capturing both the decoherence and recoherence mechanisms accurately requires considering the complexity of formation (COF), a quantity derived from the complexity of purification (COP).

This text is organised as follows: In \cref{sec:model}, we introduce the toy model of two interacting fields. In \cref{sec:Sl} and \cref{sec:COP}, we briefly review linear entropy and complexity measures, respectively. In \cref{sec:time_evol}, we discuss the time evolution of linear entropy and complexity measures. Finally, we conclude with a discussion and outline potential future directions.
\section{The Model}\label{sec:model}
We consider a toy model consisting of two interacting scalar fields: a massless field \(\varphi\) and a massive field \(\chi\) in a de Sitter (dS) background. The dS background metric is given by
\begin{align}ds^2 = a^2(\eta)\left(-d\eta^2 + d\vec{x}^2\right)\,,\end{align} where \(\eta \in (-\infty,0)\) is the conformal time, and $a(\eta)\equiv -1/(H\eta)$ is the scale factor with a constant Hubble parameter \(H\). 

Here the massless field \(\varphi\) and the heavy field \(\chi\) will, respectively, play the role of the system and of the environment. The two fields interact via a field-field coupling term \(\lambda^2 \ph\chi\), and a momentum-field coupling term \(\rho \ph'\chi\), where the coupling parameters $\lambda$ and $\rho$ both have mass dimension $1$. The prime in $\phi', \chi'$ etc will represent derivative with respect to the conformal time $\eta$ throughout the text. Consequently, the Lagrangian density for this model is given by:
\begin{equation}\label{eq:model}
\mathcal{L} =  \frac{1}{2}a^2  (\ph')^{2} - \frac{1}{2}a^2\left(\partial_i \ph\right)^2  
+ \frac{1}{2}a^2 (\chi')^{2} - \frac{1}{2}a^2 \left(\partial_i \chi\right)^2
- \frac{1}{2}M^2 a^4 \chi^2 - \lambda^2 a^4\ph\chi - \rho a^3\ph^{\prime}\chi\,.
\end{equation}
The motivation for considering this toy model arises from identifying the $\ph$ field with the adiabatic fluctuation while $\chi$ plays the role of fluctuations in one of the many heavy fields that are typically present during inflation. Indeed, from an EFT point of view, the leading order mixing term between these fields would be expressed as given above. When the inflaton fluctuations are massless (a statement that is true for the primordial curvature perturbation $\zeta$ to leading order in slow-roll), then the $\lambda$ term effectively switches off, a well-known fact for the EFT of adiabatic mode (see, for instance, \cite{Colas:2022kfu,Assassi:2013gxa}). Our Lagrangian matches theirs on considering the field redefinition  $\ph \equiv \sqrt{2\epsilon} M_{\text{Pl}}\,\zeta$. However, we keep both these terms to allow for a more general two-field model, restricted to Gaussian couplings, that is interesting from a phenomenological point of view. As it turns out, this versatility would be useful in exploring both mixedness of the state, and its ability to retain coherence, by tuning these parameters.  More explicitly, note that by turning off the momentum-field interaction (\textit{i.e.}, setting $\rho \rightarrow 0$), the model reduces to the so-called curved-space Caldeira-Leggett framework, which has been extensively studied recently in \cite{Colas2022} for demonstrating quantum decoherence. In \cite{Bhattacharyya_2024}, this approach was extended to demonstrate that quantum complexity can also serve as a useful tool for detecting decoherence.

The other end of the spectrum is when the shift symmetry of the inflaton is taken into account and the field-field coupling is turned off, as mentioned above. In this limit of the dominant momentum-field coupling, \textit{i.e.} \(\lambda \rightarrow 0\), \cref{eq:model} describes the same dynamics as the following Lagrangian density  (up to reparametrisation mentioned above) \cite{Colas:2022kfu}:
\begin{align}
\mathcal{L} &=  a^2  \epsilon M_{\mathrm{Pl}}^2 \zeta^{\prime2} - a^2 \epsilon M_{\mathrm{Pl}}^2  \left(\partial_i \zeta\right)^2  
+ \frac{1}{2}a^2 \mathcal{F}^{\prime2} - \frac{1}{2}a^2 \left(\partial_i  \mathcal{F}\right)^2
- \frac{1}{2}M^2 a^4 \mathcal{F}^2 - \rho a^3 \sqrt{2\epsilon} 
 M_{\mathrm{Pl}} \zeta^{\prime} \mathcal{F},
\end{align}
which represents a linear-order approximation of the dynamics of fluctuations in the adiabatic direction \(\zeta\) and the entropic direction \(\mathcal{F}\). Here \(\varepsilon\) is the first slow-roll parameter and \(M_{\mathrm{Pl}}\) is the Planck mass. This simple effective field theory (EFT) model was initially studied in \cite {Burgess:2017ytm,Assassi:2013gxa,Shiu:2011qw,Pinol:2020kvw,Pinol:2021aun} and more recently investigated for recoherence phenomena by \cite{Colas:2022kfu} and pointed that from an EFT perspective, $\zeta^{\prime} \mathcal{F}$ is the only operator compatible with the shift symmetry of the Goldstone mode and with spatial homogeneity of the background. This is also in keeping with the \textit{scaling} solution one finds for a two-field model of the early-universe, and focusing exclusively on inflation \cite{Tolley:2007nq}. Perturbing around the background solutions for the inflaton and the heavy field, one finds an action that is identical to \cref{eq:model} (up to field redefinitions) \cite{Colas:2024xjy}.

We introduce the Fourier modes for the fields \(\varphi\) and \(\chi\), scaled by a factor of \(a\) to write them in terms of their corresponding Mukhanov variable:
\begin{equation}\label{eq:modes}
v_\ph(\eta, \mathbf{k})\equiv a(\eta) \int_{\mathbb{R}^3} \frac{d^3\mathbf{x}}{(2\pi)^{3/2}} \ph(\eta,\mathbf{x})e^{-i\mathbf{k}\cdot\mathbf{x}},~~~\text{and}~~~v_\chi(\eta, \mathbf{k})\equiv a(\eta) \int_{\mathbb{R}^3} \frac{d^3\mathbf{x}}{(2\pi)^{3/2}} \chi(\eta,\mathbf{x})e^{-i\mathbf{k}\cdot\mathbf{x}}.
\end{equation}
In terms of these modes, the conjugate momenta are:
\begin{align}
p_\ph = v_\ph' - \frac{a'}{a}v_\ph - \rho a v_{\chi},~~~\text{and}~~~p_\chi = v_\chi' - \frac{a'}{a}v_\chi\,.
\end{align}
After performing a Legendre transform, we obtain the following Hamiltonian:
\begin{equation}\label{eq:hamiltonian}
H = \int_{\mathbb{R}^{3+}} d^3\textbf{k} \ \textbf{z}^{\dag}\textbf{H}(\eta,\textbf{k})\textbf{z}\,,
\end{equation}
where we defined
\begin{equation}
\textbf{z} \equiv \begin{pmatrix}
    v_{\varphi}\\
    p_{\varphi}\\
    v_{\chi}\\
    p_{\chi}
    \end{pmatrix} ,~~~ \textbf{H} \equiv \begin{pmatrix}
    \textbf{H}^{(\ph)} & \textbf{V}\\
    \textbf{V}^T & \textbf{H}^{(\chi)}
    \end{pmatrix}\,,
\end{equation}
with
\begin{equation}\label{Ham_expression}
\textbf{H}^{(\ph)} \equiv \begin{pmatrix}
    k^2 + m^2a^2 & \frac{a'}{a}\\
    \frac{a'}{a} & 1
    \end{pmatrix},~~~
\textbf{H}^{(\chi)} \equiv \begin{pmatrix}
    k^2 + M^2a^2 & \frac{a'}{a}\\
    \frac{a'}{a} & 1
    \end{pmatrix},~~~
\textbf{V} \equiv \begin{pmatrix}
    \lambda^2a^2 & 0\\
    \rho a & 0
    \end{pmatrix}\,.
\end{equation}

Since this model is purely Gaussian, all quantities of interest can be calculated using the covariance matrix \(\Sigma\), whose matrix elements are given by:
\begin{equation}\label{eq:cov_mat_def}
\Sigma_{ij} = \frac{1}{2}\textrm{Tr}(({\textbf{z}}_i{\textbf{z}}_j + {\textbf{z}}_j{\textbf{z}}_i)\hat{\rho}_0)\,. 
\end{equation}
Here, the density matrix $\hat{\rho}_0$ denotes the initial Bunch-Davies vacuum. Using the Heisenberg's equations of motion, we can derive the following transport equation for the covariance matrix $\Sigma$ (written in matrix form as): 
\begin{align}
    \frac{\textrm{d} \Sigma}{\textrm{d} \eta} = \Omega \textbf{H} \Sigma - \Sigma \textbf{H} \Omega \,,
\end{align}
which can be solved numerically to find the time evolution of \(\Sigma(\eta)\). Here,
\begin{equation}
\Omega \equiv \begin{pmatrix}
    \omega & 0\\
    0 & \omega
    \end{pmatrix} ,~~~ 
\omega \equiv \begin{pmatrix}
    0 & 1\\
    -1 & 0
    \end{pmatrix}\,.
\end{equation}
Of particular interest in this investigation is \(\Sigma^{(\ph\ph)}\), the top-left \(2 \times 2\) block of the \(4 \times 4\) covariance matrix \(\Sigma\). This quantifies the covariance matrix of the `system' degree of freedom (the adiabatic mode), whose diagonal elements capture the curvature and momentum power spectra, while the off-diagonal ones capture covariance between the system field and its momentum. To fix notation, we express \(\Sigma^{(\ph\ph)}\) as the following \(2\times 2\) matrix:
\begin{align}
    \Sigma^{(\ph\ph)} = \begin{pmatrix}
    \Sigma_{11} & \Sigma_{12}\\
    \Sigma_{21} & \Sigma_{22}
    \end{pmatrix} \,,
\end{align}
where $\Sigma_{21} = \Sigma_{12}$  as follows from \cref{eq:cov_mat_def}, leaving $\Sigma^{\left(\ph\ph\right)}$ with only three independent entries. Since we are interested in quantum informatic properties of the system field, as a response to tracing out the heavy entropic mode, this \(\Sigma^{(\ph\ph)}\) block of the covariance matrix will be useful for computing both the linear entropy and complexity measures. In the following sections, we provide a brief review of these quantities, followed by an analysis of their time evolution to study decoherence and recoherence.
\section{Linear Entropy}\label{sec:Sl}
Our goal is to understand the decoherence, or the lack of it (\textit{i.e.,} recoherence), for the system field given the interaction of it with the heavy field as described by the aforementioned model. As the light field interacts with the heavy field, quantum entanglement between the two builds up with time. Decoherence and, more surprisingly, recoherence can be observed by tracking the quantum state of the light field, as this interaction introduces ``mixedness" into the system. A well-known quantity for capturing this mixedness is linear entropy.

We begin with a brief overview of linear entropy, a quantity that measures the purity of a quantum system. Given a quantum system \(AB\) comprised of subsystems \(A\) and \(B\), \textit{linear entropy} (\(S_l\)) is a commonly used measure of decoherence in \(A\) due to interaction with \(B\). It is defined as follows in terms of reduced density matrix \(\hat{\rho}_A\) associated with subsystem \(A\):
\begin{align}
    S_l = 1 - \textrm{Tr}(\hat{\rho}_A^2)\,.
\end{align}
Note that here, the \textit{purity} \(\gamma \equiv \textrm{Tr}(\hat{\rho}_A^2)\) falls in the range \(1/d \leq \gamma \leq 1\), where \(d\) is the dimension of the Hilbert space. Here \(\gamma = 1\) represents a completely pure state whereas \(\gamma = 1/d\) represents a completely mixed state. In this investigation, the Hilbert space is infinite dimensional, and thus both the purity and the linear entropy lie between \(0\) and \(1\), with \(S_l = 0\) (or \(\gamma = 1\)) representing a completely pure state and \(S_l = 1\) (or \(\gamma = 0\)) representing a maximally mixed state. In terms of the covariance matrix, \(S_l\) can be calculated as \cite{Paris_2003}: 
\begin{equation}
    S_l = 1 - \frac{1}{4\det(\Sigma^{(\ph\ph)})}\,,
\end{equation}
a result that holds for a Gaussian state.
\section{Complexity Measures: COP and COF}\label{sec:COP}
Another quantity which has recently proven to be a useful measure of decoherence is circuit complexity  \cite{NL1, NL2, NL3, Jefferson,Chapman:2021jbh}. This quantity is motivated by the following question: given a target state $\ket{\psi}_T$, a reference state \(\ket{\psi}_R\) and a set of elementary gates, what is the most efficient quantum circuit which starts at \(\ket{\psi}_R\) and ends at \(\ket{\psi}_T\). This circuit is constructed from a continuous sequence of parametrized path-ordered exponentials of a control Hamiltonian operator \cite{NL1,NL2,NL3,Jefferson}. For example, the complexity \(\mathcal{C}\) for a simple harmonic oscillator is \cite{Jefferson}:
\begin{equation} \label{come}
\mathcal{C}=\frac{1}{2} \sqrt{\left(\ln\left(\frac{|\omega|}{\omega_0}\right)^2 + \arctan\left(\frac{\text{Im}(\omega)}{\text{Re}(\omega_0)}\right)^2\right)}\,,
\end{equation}
where \(\omega\) and \(\omega_0\) are the frequencies of the target state and reference state respectively. The reference state wavefunction is of the form $\psi_R(t=0,x) \sim e^{-\frac{\omega_0\, x^2}{2}}$ and $\psi_T(t,x) \sim e^{-\frac{\omega(t)\, x^2}{2}}\,,$ where $\omega(t)$ is in general complex and $\omega_0$ is real. $\psi_T(t,x)$ is obtained by time-evolving $\psi_R(t=0,x)$  with the harmonic oscillator Hamiltonian.  Also, the fundamental block for the circuit for this case is the gates generated by exponentiating scaling operator $\frac{i}{2}(\hat{x} \hat{p} + \hat{p}\hat{x})\,,$ where the hat symbol is used to denote the operators and $[\hat{x},\hat{p}]=i\,.$ For more details readers are referred to Sec.~(3.2) of \cite{Ali:2018fcz}. Furthermore, one can easily check that, when $\omega \rightarrow\omega_0,$ the complexity defined in  eq.~(\ref{come}), vanishes as the target and reference coincide in this limit.

However, the method described above is for a pure state by construction. Extending the computation of complexity to open quantum systems is still an active area of research, however the \textit{complexity of purification} (COP) \cite{Caceres:2019pgf,Agon:2018zso,Camargo:2018eof, Ghodrati:2019hnn,Bhattacharyya:2020iic} is one such extension. What follows is a description of how COP is calculated.

Given a Hilbert space \(\mathcal{H}\) and a mixed state \(\hat{\rho}_{\textrm{mix}}\), we can purify \(\hat{\rho}_{\textrm{mix}}\) to a pure state \(\ket{\psi}\) in an enlarged Hilbert space \(\mathcal{H}_{\textrm{pure}} = \mathcal{H} \otimes \mathcal{H}_{\textrm{anc}}\) where \(\mathcal{H}_{\textrm{anc}}\) represents ancillary degrees of freedom. We also impose the restriction,
\begin{align}
    \textrm{Tr}_{\mathcal{H}_{\textrm{anc}}}(\ket{\psi}\bra{\psi}) = \hat{\rho}_{\textrm{mix}} \,,\label{eq:rho_mix_constraint}
\end{align}
in order to ensure that the original state is obtained when tracing out the ancillary Hilbert space. This constraint guarantees that any observable in the original Hilbert space $\mathcal{H}$ retains its expectation value under the purification:
\begin{align}
    \langle \hat{\mathcal{O}} \rangle = \textrm{Tr}_{\mathcal{H}}(\hat{\rho}_{\textrm{mix}} \hat{\mathcal{O}}) = \textrm{Tr}_{\mathcal{H}_{\textrm{anc}}}(\bra{\psi} \hat{\mathcal{O}} \otimes \mathcal{I}_{\textrm{anc}} \ket{\psi})\,.
\end{align}
Given that the choice of ancillary Hilbert space is not unique (and hence the purification process is dependent on this choice), a set of pure states \(\ket{\Psi}_{\alpha, \beta, \gamma,...}\) can be constructed, parametrised by \(\alpha, \beta, \gamma,...\) which satisfy the purification requirement. We then minimize the quantity of interest (in this case, the complexity $\mathcal{C}$) with respect to these parameters. Thus, the COP is obtained as \cite{Bhattacharyya:2020iic}:
\begin{align}
    \textrm{COP} \equiv \min_{\alpha, \beta, \gamma, ...} \mathcal{C}(\ket{\Psi}_{\alpha, \beta, \gamma,...}, \ket{\Psi}_R)\,,
\end{align}
where \(\ket{\Psi}_R\) is the reference state and \(\mathcal{C}\) is the complexity as calculated in the enlarged Hilbert space. Following \cite{Caceres:2019pgf, Bhattacharyya:2018sbw, Bhattacharyya:2020iic}, we choose to use minimal purification such that the size  system  is same as the ancillary system as well as restrict ourselves to Gaussian purification\footnote{It will be interesting to go beyond the minimal purification scheme. We leave this for future studies.}. Given these, the purified state can be parametrized as
\begin{align} \label{pure}
    \psi(\rm{z},\rm{z}_{\rm{anc}}) \sim \exp{\left\{-\frac{1}{2}\left(\alpha\, \rm{z}^2 + \beta\, \rm{z}_{\rm{anc}}^2 - 2\tau \,\rm{z}\, \rm{z_{anc}} \right)\right\}}\,,
\end{align}
where $\rm{z}$ and $\rm{z_{anc}}$ belong to the Hilbert spaces $\mathcal{H}$ and $\mathcal{H}_{\rm anc}$ respectively. The parameters $\alpha, \beta$ and $\tau$ can be specified in the following way: trace out the ancillary degrees of freedom of the density matrix $\ket{\psi}\bra{\psi}$ to find the reduced density matrix $\hat{\rho}_{\rm{mix}}$ which also has a Gaussian structure and hence can readily be linked to the covariance matrix $\Sigma^{(\ph\ph)}$. Then following \cite{Bhattacharyya:2020iic}, this fixes $\alpha, \tau $ and $\text{Re}(\beta)\,.$

Now notice that the form of $\psi(\rm{z},\rm{z}_{\rm{anc}})$ is that of two-coupled oscillators. Then, a straightforward generalization of eq.~(\ref{come}) for the two-oscillator case\footnote{Readers are referred to \cite{Ali:2018fcz} for more details. \cite{Ali:2018fcz} contains an expression for the circuit complexity for N-coupled oscillators in terms of the normal-mode frequencies. The expression under the square root in eq.~(\ref{COP}) is just a special case of the expression quoted in eq.~(3.44) of \cite{Ali:2018fcz}.} gives an expression for the COP for our model as \cite{Bhattacharyya:2020iic}:
\begin{align}\label{COP}
    \rm{COP} = \min_{\text{Im}(\beta)}\frac{1}{2}\sqrt{\sum_{i = 1}^2\left(\ln\left(\frac{|\omega_i|}{\omega_0}\right)^2 + \arctan\left(\frac{\text{Im}(\omega_i)}{\text{Re}(\omega_0)}\right)^2\right)}\,,
\end{align}
where
\begin{eqnarray}
    \omega_1 &\equiv& \frac{1}{2}\left(\alpha + \beta + \sqrt{(\alpha - \beta)^2 + 4\tau^2}\right)\,,\\
    \omega_2 &\equiv& \frac{1}{2}\left(\alpha + \beta - \sqrt{(\alpha - \beta)^2 + 4\tau^2}\right)\,,
\end{eqnarray}
and
\begin{equation} \label{fixp}
    \alpha = \Sigma^{(\ph\ph)}_{11}, ~~~ \text{Re}(\beta) = 2\Sigma^{(\ph\ph)}_{22}, ~~~ \tau^2 = 4\left(\Sigma^{(\ph\ph)}_{12}\right)^2 - 1 + 4i\Sigma^{(\ph\ph)}_{12}\,.
\end{equation}
The frequency $\omega_0$ of the reference state is set to 1 for convenience. $\omega_1$ and $\omega_2$ are the normal-mode frequencies in the purified space. Note that eq.~(\ref{fixp}) fixes all the parameters except \(\text{Im}(\beta)\) for a given $\Sigma^{(\ph\ph)}\,.$ Thus \(\text{Im}(\beta)\) is treated as a free parameter, and in order to compute COP, the expression in \eqref{COP} is minimised over \(\text{Im}(\beta)\).

The complexity of scalar fields in de Sitter, especially that of light or massless ones, has been shown to increase due to the squeezing associated with the background spacetime \cite{Bhattacharyya:2020rpy, Bhattacharyya:2020kgu}. The reasoning behind this is that, in this computation, one assumes the reference state  to be the Fock vacuum for short wavelength modes deep inside the horizon, while the target state is the squeezed one once the mode exits the horizon. As the wavelength of the mode gets more and more redshifted, the squeezing parameter increases and thus the complexity rises unboundedly.\footnote{This, of course, is only true for an idealized de Sitter space and not if inflation ends at some finite time followed by a subsequent radiation era.} However, in order to see features in the COP that can be identified with recoherence, we need to isolate them from the complexity due to squeezing. Put another way, recoherence happens due to the effect of the interaction between the two fields, and we need to subtract the effect the gravitational squeezing has on the complexity to see this clearly. Keeping this in mind, we define a complexity measure called the \textit{complexity of formation} (COF) \cite{Chapman:2018hou}, which is found by subtracting the COP of the free theory from that of the interacting theory as follows: 
\begin{align}
    \rm{COF} \equiv \left\vert\,\rm{COP} - \rm{COP}_{\lambda = 0,\,\rho = 0}\,\right\vert\,,
\end{align}
where \(\rm{COP}_{\lambda = 0,\, \rho = 0}\) is the COP associated with zero coupling between the fields \(\ph\) and \(\chi\).
\section{Time evolution of Linear Entropy, COP and COF}\label{sec:time_evol}
In this section, we will first discuss the time evolution of linear entropy, COP and COF with either the field-field coupling $\lambda$ or the momentum-field coupling $\rho$ enabled. Next, we will investigate the case of both types of couplings enabled and show the correspondence between the qualitative features of the time evolution of the linear entropy and the complexity.
\subsection{Field-Field Coupling $\lambda$ vs Momentum-Field Coupling $\rho$}\label{sec:one_coupling}
\begin{figure}
    \begin{subfigure}[t]{0.5\linewidth}
        \centering
        \includegraphics[width=\linewidth]{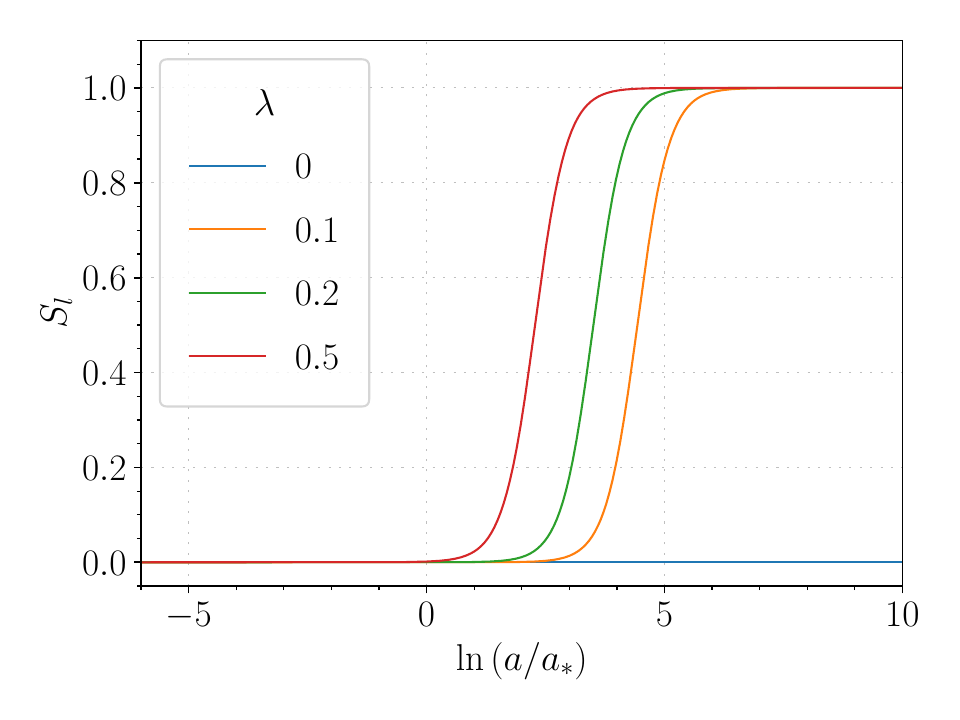} 
        \caption{Field-field coupling} \label{Fig:Sl_no_rho}
    \end{subfigure}
    \begin{subfigure}[t]{0.5\linewidth}
        \includegraphics[width=\linewidth]{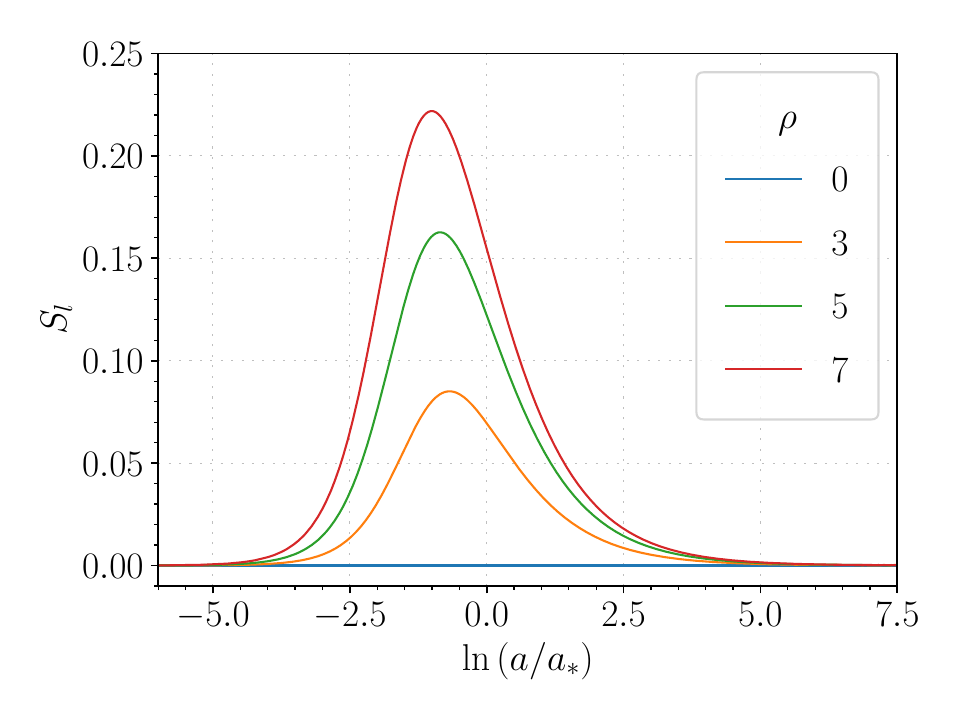} 
        \caption{Momentum-field coupling} \label{Fig:Sl_no_lambda}
    \end{subfigure}
    \caption{Time evolution of linear entropy \((S_l)\) in the special case where either the field-field coupling (Fig.~(\ref{Fig:Sl_no_rho})) or the momentum-field coupling (Fig.~(\ref{Fig:Sl_no_lambda}))  is enabled. Here, \(H = 1\), \(M = 4\) and \(k = 1\) in both plots.}\label{Fig:Sl_two_cases}
\end{figure}

In the case where we only have one type of coupling between \(\ph\) and \(\chi\) in \cref{eq:model} (\textit{i.e.,} either \(\rho = 0\) or \(\lambda = 0\)), the time evolution of linear entropy is well-understood \cite{Colas2022,Colas:2021llj,Bhattacharyya_2024}. Nevertheless, we review these results here for the sake of comparison with the time evolution of complexity. Fig. \ref{Fig:Sl_two_cases} displays the comparison between the two cases for various values of the couplings. We shall always be tracking the behavior of a given Fourier mode in our analyses since the underlying system is a Gaussian one with no mode-coupling involved. 
We immediately notice that the late-time behavior of the linear entropy is qualitatively different for these two cases. In the case with only the field-field coupling (Fig.~(\ref{Fig:Sl_no_rho})), the linear entropy follows a sigmoid curve, suddenly increasing at a timescale inversely proportional to the field-field coupling \(\lambda\) and saturates near 1. This suggests that at late times the system has decohered completely. On the other hand, in the case with only the momentum-field coupling (Fig.~(\ref{Fig:Sl_no_lambda})), the linear entropy undergoes a transient period of increase after which it decreases and settles back near 0 (on choosing a region of parameter space such that the entropic field is sufficiently heavy, which is the case for all the cases considered in Fig.~(\ref{Fig:Sl_no_lambda})). This suggests that the system transiently becomes more mixed, and then recoheres to remain pure at late times. The timescale at which the system starts to recohere is independent of the momentum-field coupling strength $\rho$. It is important to note that while the degree to which decoherence takes before recoherence increases with $\rho$ for the momentum-field coupling, it certainly does not lead to complete decoherence as observed in the field-field coupling case. This feature highlights the role of the specific interaction terms: the momentum-field coupling causes partial decoherence before the system recoheres, in contrast to the case of the field-field coupling which induces complete decoherence.

The physical reason behind the recoherence is that the interaction between the system and environment `switches off' dynamically due to effects of the curved background, allowing the coherence of the state to freeze at some value after information backflows from the environment to the system. Such oscillations in the linear entropy of a system is only possible when the system mode couples to a very few degrees of freedom of the environment (here ${\bf k}$ modes can only couple to $-{\bf k}$ modes due to momentum-conservation), and is a classic signature of non-Markovian behaviour.

\begin{figure}
    \begin{subfigure}[t]{0.50\textwidth}
        \centering
\includegraphics[width=\linewidth]{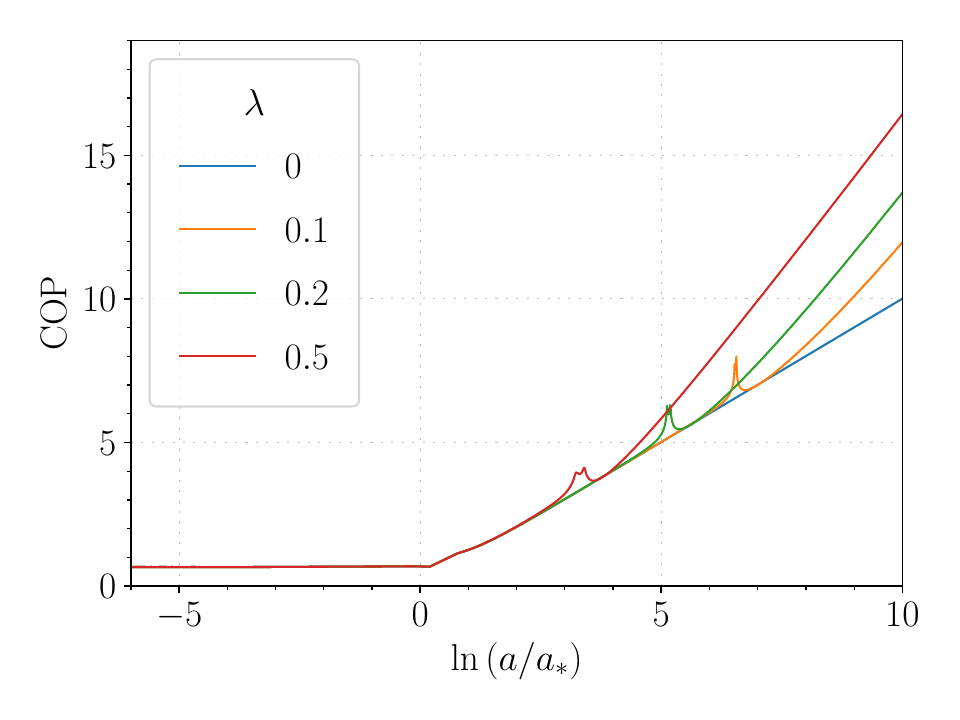} 
        \caption{Field-field coupling} \label{Fig:COP_no_rho}
    \end{subfigure}
    \begin{subfigure}[t]{0.50\textwidth}
        \includegraphics[width=\linewidth]{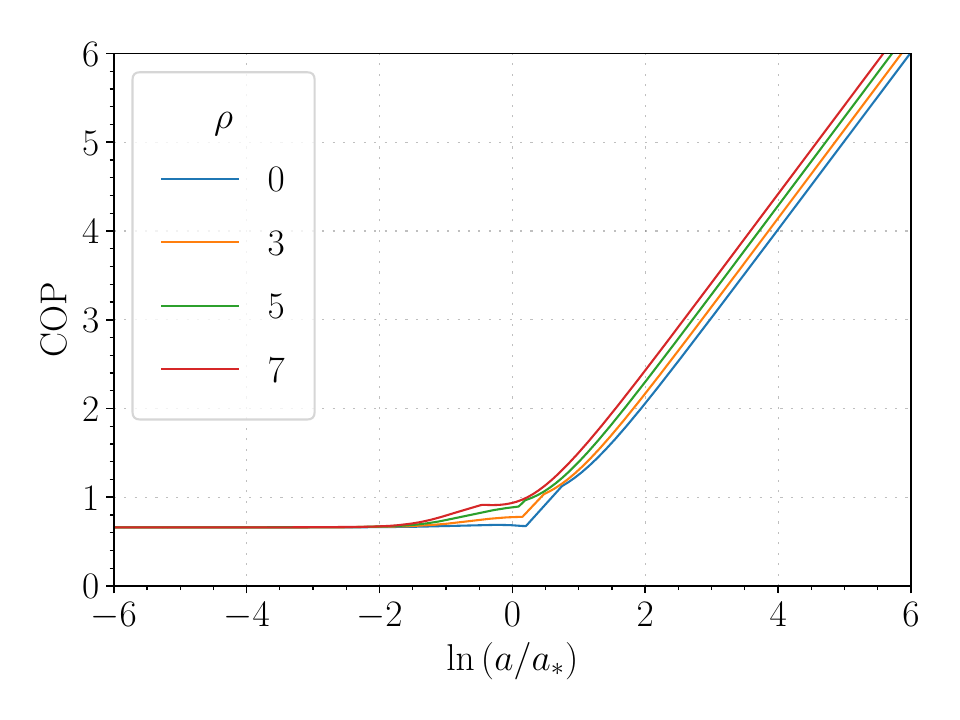} 
        \caption{ Momentum-field coupling} \label{Fig:COP_no_lambda}
    \end{subfigure}
    \caption{Time evolution of complexity of purification (COP) in the special case where either the field-field coupling  (Fig.~(\ref{Fig:COP_no_rho})) or the momentum-field coupling(Fig.~(\ref{Fig:COP_no_lambda})) is enabled. Here, \(H = 1\), \(M = 4\) and \(k = 1\) in both plots.}\label{Fig:COP_two_cases}
\end{figure}
\begin{figure}[t]
    \centering
    \begin{subfigure}[t]{0.49\textwidth}
        \centering
        \includegraphics[width=\linewidth]{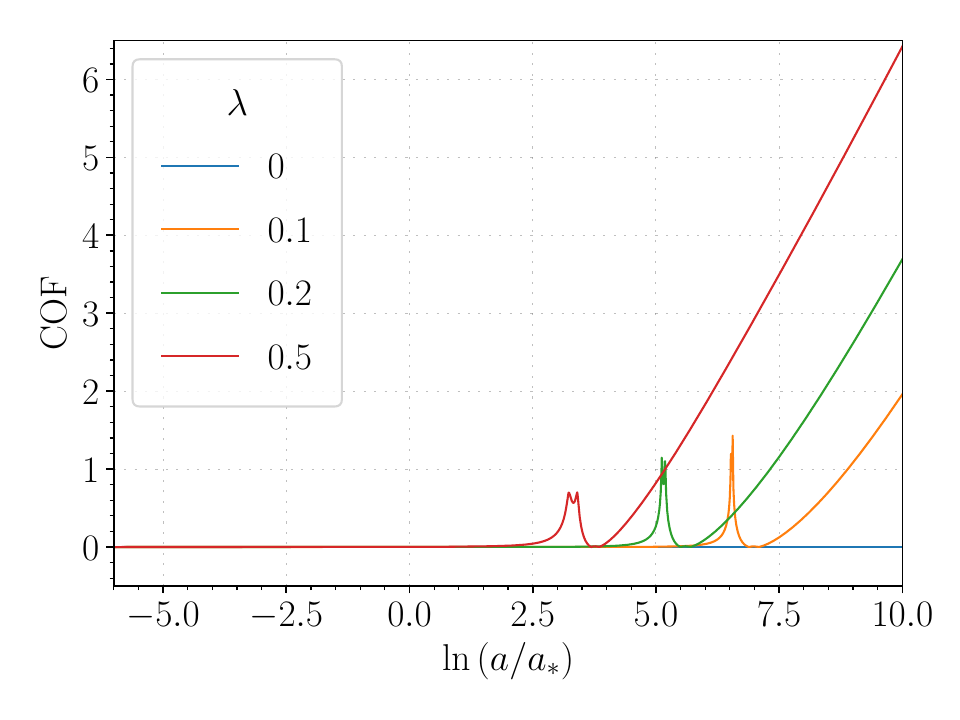} 
        \caption{Field-field coupling} \label{Fig:COF_no_rho}
    \end{subfigure}
    \begin{subfigure}[t]{0.49\textwidth}
        \centering
        \includegraphics[width=\linewidth]{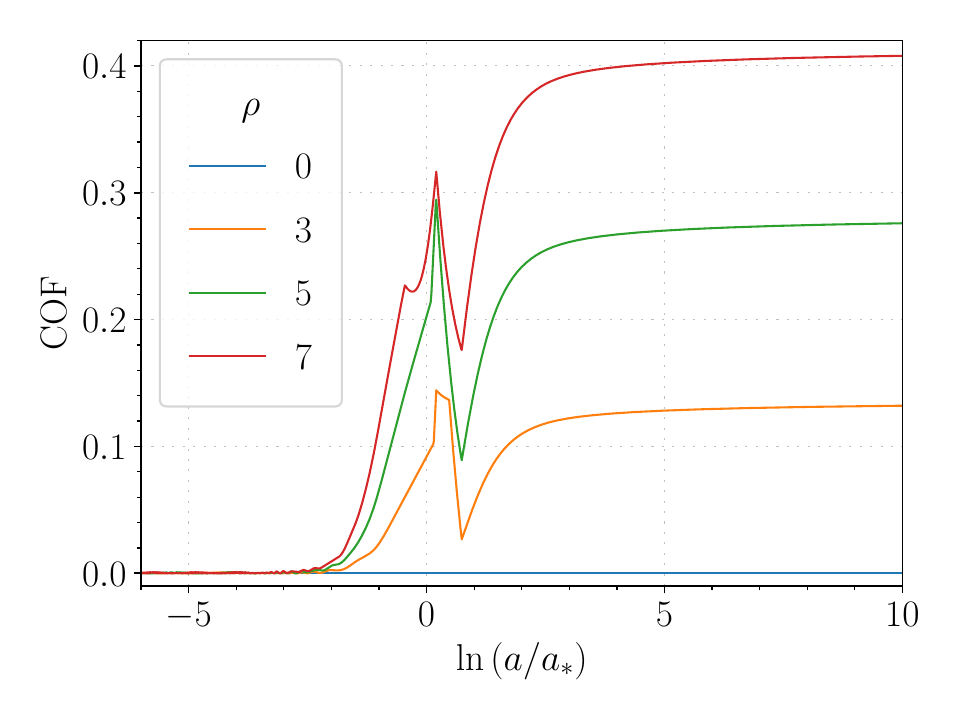} 
        \caption{ Momentum-field coupling} \label{Fig:COF_no_lambda}
    \end{subfigure}
    \caption{Time evolution of complexity of formation (COF) in the special case where either the field-field coupling (Fig.~(\ref{Fig:COF_no_rho})) or the momentum-field coupling (Fig.~(\ref{Fig:COF_no_lambda})) is enabled. Here, \(H = 1\), \(M = 4\) and \(k = 1\) in both plots.}\label{Fig:COF_two_cases}
\end{figure}

Now we will compare the above behavior with that of the measures of complexity, namely COP and COF. With only the field-field coupling enabled (Fig.~\ref{Fig:COP_no_rho}), the COP is initially fairly close to 0, followed by a short period of intermediate growth, and finally a late-time rapid and unbounded growth. The intermediate growth and late-time growth periods are separated by a sudden bump. The late-time growth starts at a timescale similar to the saturation timescale of linear entropy. Moreover, the timescale of this COP growth is inversely proportional to the coupling strength $\lambda$. This is indicative of the fact that COP is sensitive to decoherence for the case of field-field coupling (Fig.~(\ref{Fig:COP_no_rho})) \cite{Bhattacharyya_2024}. The time evolution of COF also shows a similar feature as the late-time growth of COP (see Fig.~(\ref{Fig:COF_no_rho})) for the field-field coupling, hence a similar correspondence can be suggested between the COF growth and the decoherence of the system. 

Interestingly, when only the momentum-field coupling is enabled, the time evolution of COP does not show any clear feature that is indicative of recoherence (see Fig.~(\ref{Fig:COP_no_lambda})). We suspect that this is due to the large squeezing effect as mentioned in the earlier section. The background squeezing obfuscates any signature of the decrasing linear entropy in the COP and one needs to separate the two to find any discernable signal. In this regard, COF will be a useful tool to probe the recoherence of the system. As demonstrated in Fig.~(\ref{Fig:COF_no_lambda}), COF shows a sudden
increase followed by a rapid decrease during the transient period of decoherence to recoherence, and finally saturates rather than linear growth.

This saturation is a distinctive feature of COF and to understand its implication we will compare its timescale with that of the features observed in the time evolution of linear entropy. The time evolution of linear entropy shown in Fig.~(\ref{Fig:Sl_no_lambda}) suggests that the saturation of COF signifies the completion of recoherence of the system. In addition, we notice that the saturation value of COF increases with the coupling strength $\rho$. This is similar to the case of linear entropy where the peak before recoherence increases with increasing coupling strength.

In summary, the comparison between the qualitative features of the time evolution of complexity measures and linear entropy suggests the following:
\begin{itemize} 
\item The time evolution of COF in a de Sitter background exhibits distinctive features: field-field coupling leads to complete decoherence, whereas momentum-field coupling results in recoherence. In contrast, looking at the COP alone does not give any distinctive signature of recoherence, as COP behaves in a similar manner in both cases.
   \item The linear growth of COF signals the decoherence of the system. In other words, if the system approaches to a completely mixed state, the COF starts growing linearly and unboundedly.
    \item If the system has recohered, \textit{i.e.,} goes back to a pure state, the COF will asymptotically saturate to a constant value. 
    \item The timescale for decoherence depends on the coupling, whereas the timescale of recoherence is insensitive to coupling. This fact is also reflected from the behaviour of the COF.
\end{itemize} 
In the following subsection we will turn on both the field-field and momentum-field couplings and compare the features of the time evolution of linear entropy and COF.

\subsection{Both Field-Field and Momentum-Field Coupling Enabled}

\begin{figure}[htb!]
    \begin{subfigure}[t]{0.5\textwidth}
        \centering
        \includegraphics[width=\linewidth]{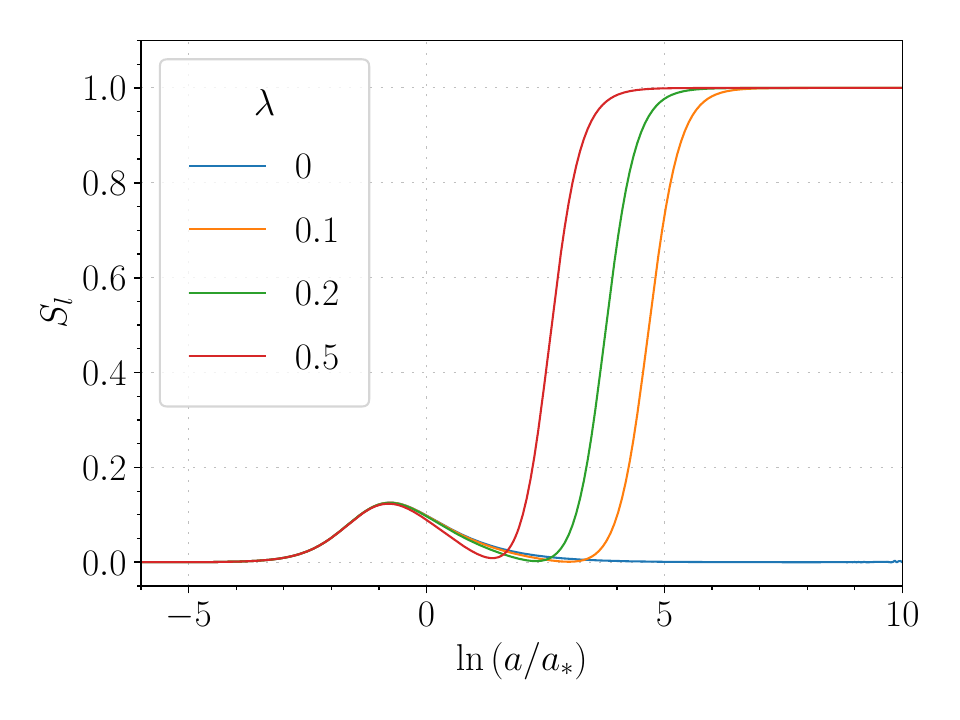} 
        \caption{$S_l$} \label{Fig:S_l_both_couplings_diff_lambda}
    \end{subfigure}
    \begin{subfigure}[t]{0.5\textwidth}
        \centering        \includegraphics[width=\linewidth]{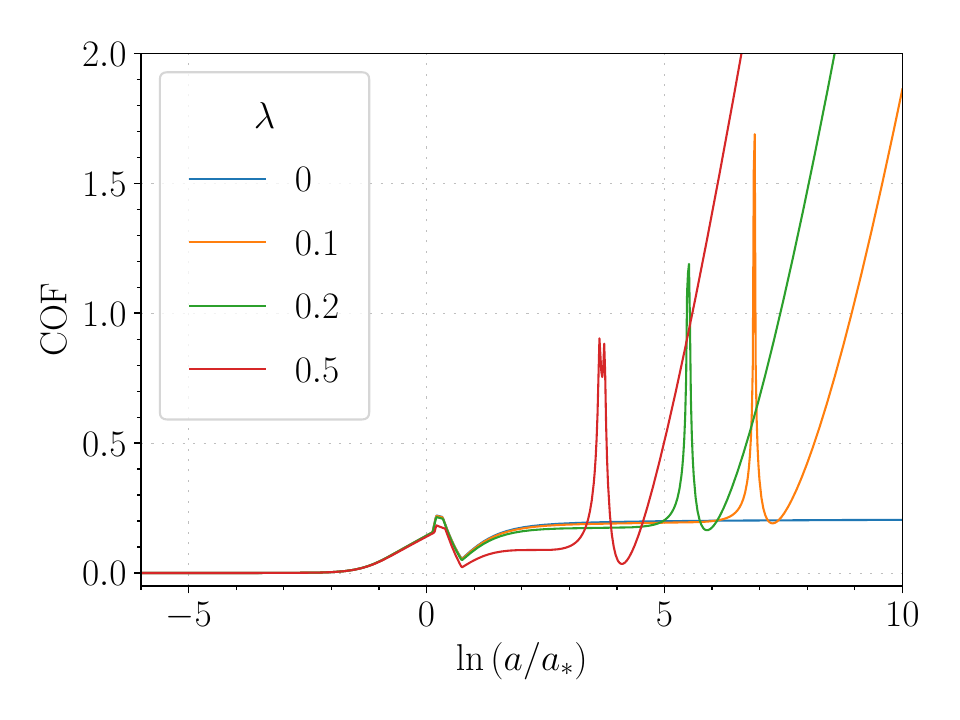} 
        \caption{COF} \label{Fig:COF_both_couplings_diff_lambda}
    \end{subfigure}
    \caption{Time evolution of linear entropy \((S_l)\) and complexity of formation (COF) for varying field-field coupling $\lambda$ with a fixed momentum-field coupling \(\rho = 4\). Here, \(H = 1\), \(M = 4\) and \(k = 1\) in both plots.}\label{Fig:both_couplings_diff_lambda}
\end{figure}

\begin{figure}[htb!]
    \begin{subfigure}[t]{0.5\textwidth}
        \centering
        \includegraphics[width=\linewidth]{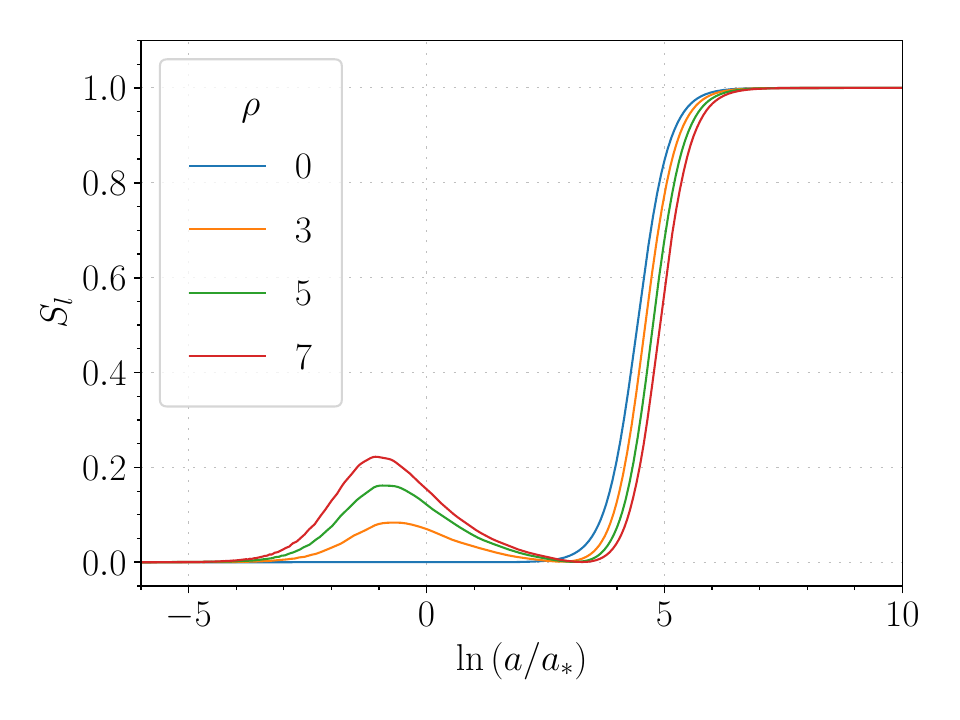} 
        \caption{$S_l$} \label{Fig:S_l_both_couplings}
    \end{subfigure}
    \begin{subfigure}[t]{0.5\textwidth}
        \centering
        \includegraphics[width=\linewidth]{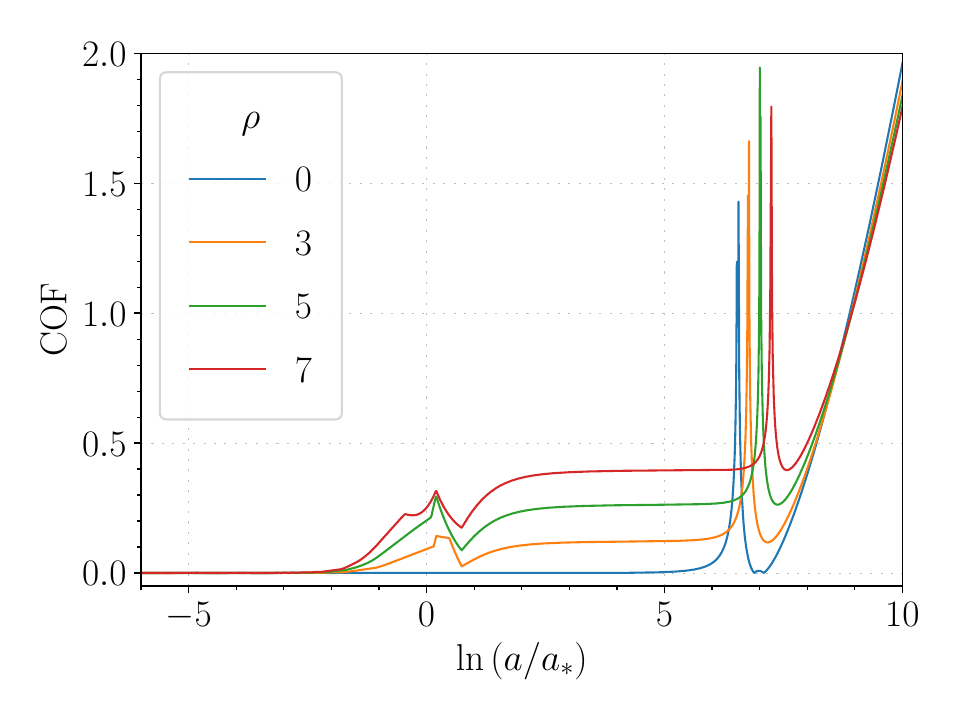} 
        \caption{COF} \label{Fig:COF_both_couplings}
    \end{subfigure}
    \caption{Time evolution of linear entropy \((S_l)\) and complexity of formation (COF) for varying momentum-field coupling $\rho$ with a fixed  field-field coupling  \(\lambda = 0.1\,\). Here, \(H = 1\), \(M = 4\) and \(k = 1\) in both plots.}\label{Fig:both_couplings}
\end{figure}

When we turn on both the field-field and the momentum-field couplings, we observe a superposition of the characteristics of the individual couplings discussed in the previous subsection. For example, linear entropy shows an early recoherent phase, followed by a late-time saturation indicating a transition to a maximally-mixed state (see Fig.~(\ref{Fig:S_l_both_couplings_diff_lambda}) and Fig.~(\ref{Fig:S_l_both_couplings})). One can see that the COF faithfully reproduces this behaviour. It shows a bump at early times -- followed by a plateau -- which signifies the recoherence. At late-times, the time evolution of COF shows a sharp kink, followed by a linear growth, which signals decoherence (see Fig.~(\ref{Fig:COF_both_couplings_diff_lambda}) and Fig.~(\ref{Fig:COF_both_couplings})). 

\begin{figure}[t]
    \begin{subfigure}[t]{0.5\textwidth}
        \centering
        \includegraphics[width=\linewidth]{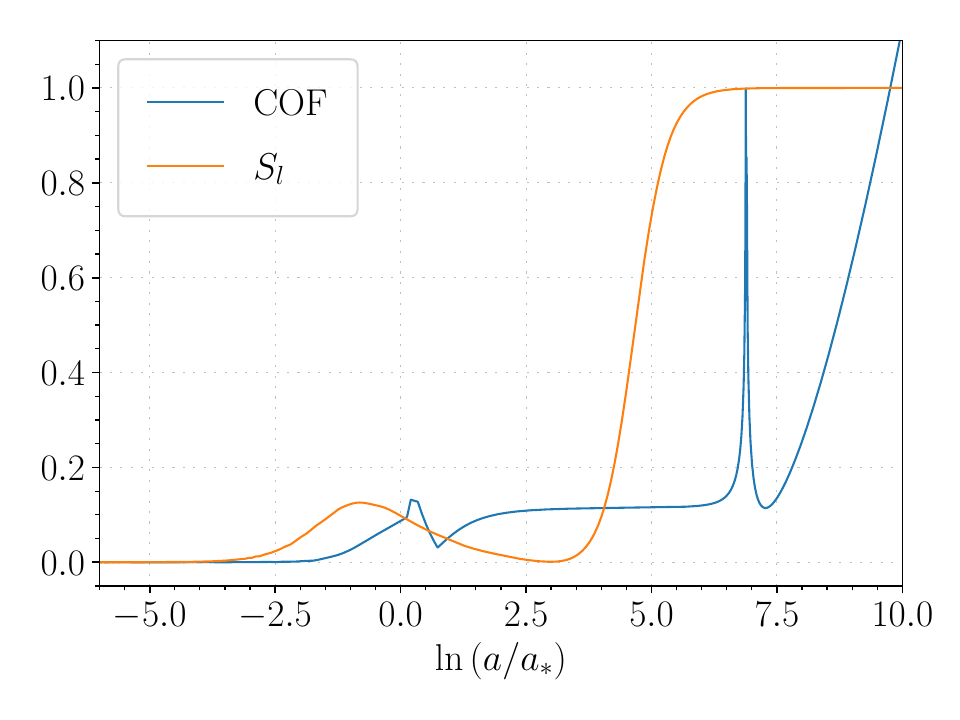} 
        \caption{$\lambda = 0.1$} \label{Fig:comp_1}
    \end{subfigure}
    \begin{subfigure}[t]{0.5\textwidth}
        \centering
        \includegraphics[width=\linewidth]{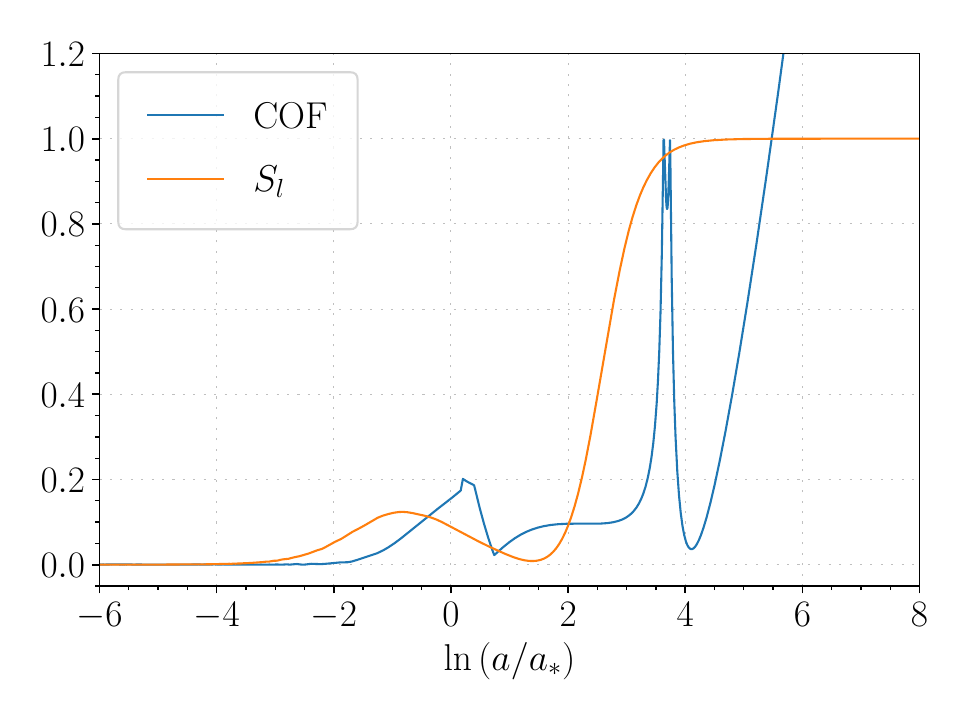} 
        \caption{$\lambda = 0.5$} \label{Fig:comp_3}
    \end{subfigure}
    \caption{Comparison of the time evolution of linear entropy \((S_l)\) and complexity of formation (COF). Note that in these plots, COF has been normalised to the same scale as $S_l$ for the ease of comparison.  Here, \(H = 1\), \(M = 4\), \(\rho = 4\) and \(k = 1\) in both plots.}\label{Fig:COFvsSl}
\end{figure}

\begin{figure}[htb!]
    \begin{subfigure}[t]{0.50\textwidth}
        \centering
        \includegraphics[width=\linewidth]{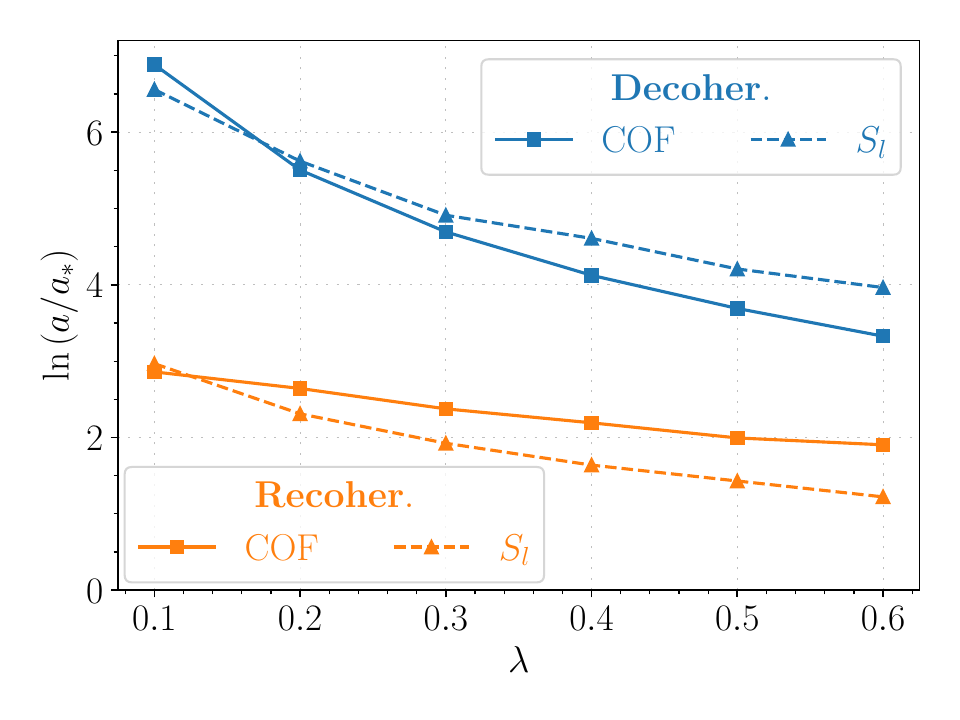} 
        \caption{$\rho = 4$} \label{Fig:time-scales_vary_lambda}
    \end{subfigure}
    \begin{subfigure}[t]{0.50\textwidth}
        \centering
        \includegraphics[width=\linewidth]{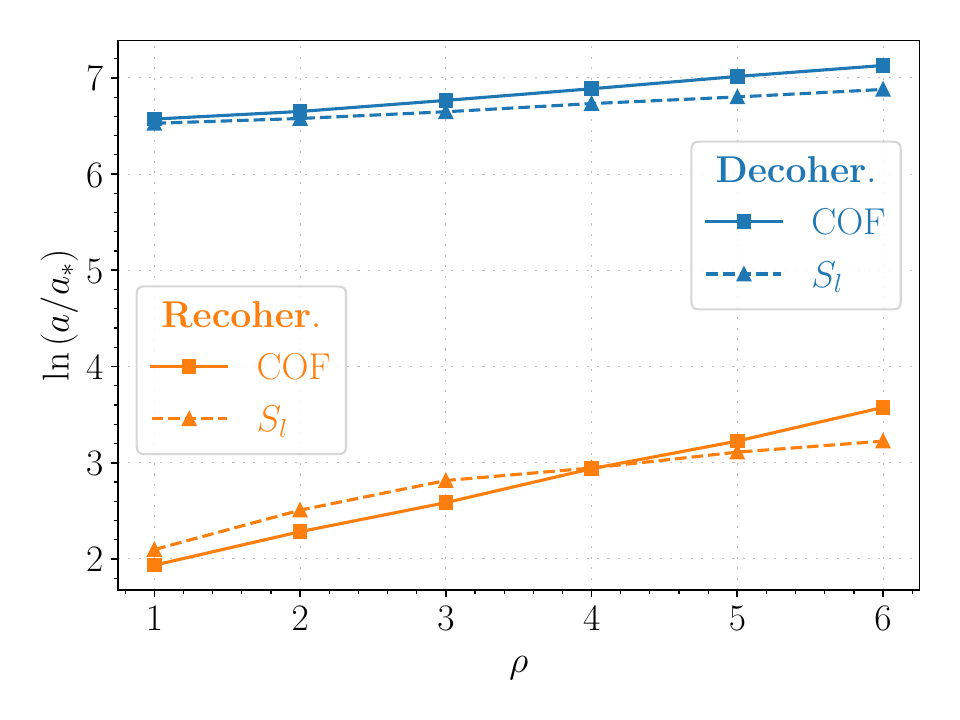} 
        \caption{$\lambda = 0.1$} \label{Fig:time-scales_vary_rho}
    \end{subfigure}
    \caption{The blue (orange) plots compare the decoherence (recoherence) timescales of linear entropy \( (S_l) \) (dashed lines) and complexity of formation (COF) (solid lines). In the left (right) panel, the coupling parameter \( \lambda \) (\( \rho \)) is varied while keeping the other coupling fixed at \( \rho = 4 \) (\( \lambda = 0.1 \)). The other parameters are set as \( H = 1 \), \( M = 4 \), and \( k = 1 \) in both panels.
}\label{Fig:time-scales}
\end{figure}

At this point, it is important to compare the timescales of these above characteristic features corresponding to the recoherence and the saturation of decoherence for the linear entropy and the COF time evolution---previously illustrated separately in the left and right panels of Figs.~(\ref{Fig:both_couplings_diff_lambda}) and (\ref{Fig:both_couplings}). Figure~(\ref{Fig:COFvsSl}) overlays the dynamics of both quantities for two representative parameter choices, clearly demonstrating the alignment of their characteristic timescales. In addition, over a range of coupling strength parameters $\lambda$ and $\rho$, the timescales of these characteristic features from the linear entropy and the COF evolution are compared in Fig.~(\ref{Fig:time-scales}) further revealing a clear similarity.  This reinforces the interpretation of these features as signatures of recoherence and decoherence in the COF dynamics suggested in Sec.~(\ref{sec:one_coupling}). More precisely, the COF recoherence timescale---plotted in Fig.~(\ref{Fig:time-scales})---is identified at the point where the COF temporarily ``flattens out'' and remains roughly constant before the later phase of decoherence. Similarly, the recoherence timescale in linear entropy is marked by the point at which the linear entropy returns to 0 after the initial temporary phase of decoherence. Finally, the decoherence timescales are characterized by the late-time ``bump" in COF (as discussed in \cite{Bhattacharyya_2024}) and the point at which the linear entropy is sufficiently close to saturation at 1.\footnote{Since the linear entropy never reaches exactly 1, we chose 0.998 as a ``sufficiently close" benchmark.}

The physical reason behind the dynamical behaviour of the linear entropy, and the COF, can be understood as follows. As long as the field-field coupling $\lambda$ is not zero, it always dominates at late-times and leads to complete decoherence of the system quantum field. This can be understood by examining the strength of the field-field coupling as compared to that of the field-momentum coupling $\rho$. Firstly, the former is accompanied by a factor of $a^2$ while the latter by that of $a$, and recall that the scale factor grows exponentially during inflation. More importantly, the form of the operators corresponding to the two terms is such that the $\zeta' \mathcal{F}$ term is proportional to the decaying mode of the adiabatic perturbation, which decays very fast after horizon crossing. This is the main reason why the field-field coupling term wins over the field-momentum coupling term (see \cref{Ham_expression}) if one waits long enough. The second feature to note is that the linear entropy, corresponding to smaller momentum-field coupling, crosses over that for a larger value of $\rho$ during this late-time phase once decoherence takes over\footnote{From Fig.~(\ref{Fig:S_l_both_couplings}), this can be seen as the orange curve crosses over the green and red curves at late-times.}. A similar behavior can be seen for the COFs of the corresponding plots in Fig.~(\ref{Fig:both_couplings}). This is because the faster decoherence (or growth in $S_l$ or COF) will take place for the smaller value of $\rho$ as this term tries to slow down this behavior.  In addition, we notice for a fixed $\rho$, the stronger field-field coupling $\lambda$ shifts the timescale of decoherence to an earlier time, whereas the recoherence timescale remains constant (see Fig.~(\ref{Fig:both_couplings_diff_lambda})).

We conclude this section with a remark on the coupling strength considered in this paper. In a Gaussian theory, the coupling strength $\lambda$ can be chosen freely.  For the purposes of this paper, we restrict our analysis to the parameter space $\lambda \ll M$. This, in essence, is required to examine the part of the parameter space that allows for recoherence. In other words, if the heavy field is not sufficiently massive, its decaying mode fails to quench the momentum-field interaction term, preventing recoherence.

\section{Discussion}
In this work, we explored the interplay between decoherence and recoherence in cosmological systems using a toy model of interacting scalar fields in a de Sitter background. By focusing on the dynamics of linear entropy, complexity of purification (COP), and complexity of formation (COF), we provided a nuanced understanding of how these measures evolve under varying coupling conditions and highlighted their relevance in capturing both qualitative and quantitative features of the system. Our findings underscore the intricate relationships between coupling types, decoherence, and the potential for recoherence in quantum systems.

The study revealed distinct timescales associated with decoherence and recoherence. In the case of field-field coupling, the system undergoes complete decoherence, with linear entropy following a characteristic sigmoid-like growth and saturating at late times. On the other hand, momentum-field coupling induces a transient period of decoherence, after which the system recoheres and stabilizes in a coherent state. An intriguing aspect of this behavior is that the recoherence timescale remains independent of the coupling strength, whereas the timescale for decoherence varies (inversely) in proportion to the coupling. These findings demonstrate how the interplay between coupling dynamics and decoherence-recoherence transitions shapes the overall evolution of the system.

Our results further highlight the role of coupling type in influencing complexity measures such as COP and COF. For field-field coupling, both COP and COF exhibit unbounded late-time growth directly reflecting the system's decoherence. By contrast, for momentum-field coupling, COF saturates at late times, signaling the completion of recoherence, with its peak value scaling with the coupling strength. This correspondence between complexity measures and linear entropy demonstrates the complementary nature of these diagnostics and their ability to capture the intricate behavior of quantum systems.

When both field-field and momentum-field couplings are simultaneously enabled, the system displays a superposition of features associated with the individual couplings. Linear entropy shows an early recoherence followed by late-time saturation, while COF captures multiple transitions, including an early-time bump -- followed by a temporary saturation -- indicative of recoherence and a sharp late-time kink marking decoherence. Comparing the timescales of linear entropy and COF dynamics strengthens the case for COF as an alternative diagnostic tool for examining the interplay between coherence and decoherence in open quantum systems.

Our findings based on complexity might be useful for theoretical model building for understanding cosmology. The complementary roles of COP and COF emphasize the importance of coupling-specific features in shaping the decoherence and recoherence dynamics of quantum systems. Our study provides a foundation for exploring broader implications of quantum recoherence in the early universe, including connections to quantum information theory and the emergence of classicality in cosmological settings.

Looking ahead, this study opens several promising avenues for future research. Refining diagnostic tools to develop more sophisticated complexity measures that account for both squeezing effects and coupling-specific dynamics could enhance our understanding of recoherence phenomena. Extending the analysis to more realistic cosmological scenarios, such as radiation- and matter-dominated epochs following a phase of quasi-de Sitter expansion, may provide a more comprehensive picture of decoherence-recoherence transitions. More interestingly, one needs to adapt a suitable definition of circuit complexity, as applicable for non-Gaussian systems, to include the cubic and higher order mode-couplings in the action mixing the curvature perturbation with the entropic sector. This is essential since General Relativity is non-linear and it is crucial to see if the features found in the dynamics of complexity can survive the addition of perturbative non-Gaussianities. In an independent direction, the theoretical insights presented here could inform experimental efforts to study open quantum systems, particularly in analog models of cosmology or quantum simulations.

In conclusion, this work underscores the intricate interplay between complexity, decoherence and recoherence in quantum cosmology. By establishing COF as a versatile and sensitive diagnostic tool, we provide a pathway for probing the fundamental dynamics of quantum systems in the early universe, with potential implications for both foundational physics and quantum information theory.

\section*{Acknowledgment}
 AB is supported by the Core Research Grant (CRG/2023/ 001120) and Mathematical Research Impact Centric Support Grant (MTR/2021/000490)  by the Department of Science and Technology Science and Anusandhan National Research Foundation (formerly SERB), Government of India. AB also acknowledges the associateship program of the Indian Academy of Sciences (IASc), Bengaluru. SB is supported in part by the Higgs Fellowship and by the STFC Consolidated Grant ``Particle Physics at the Higgs Centre''. SSH is supported in part by the “Quantum Technologies for Sustainable Development” grant from the National Institute for Theoretical and Computational Sciences of South Africa (NITHECS).

\bibliographystyle{JHEP}
\bibliography{biblio}

\providecommand{\href}[2]{#2}\begingroup\raggedright\begin{thebibliography}{10}

\bibitem{Lehners:2008vx}
J.-L.~Lehners, \emph{{Ekpyrotic and Cyclic Cosmology}},
  \href{https://doi.org/10.1016/j.physrep.2008.06.001}{\emph{Phys. Rept.}
  {\bfseries 465} (2008) 223}
  [\href{https://arxiv.org/abs/0806.1245}{{\ttfamily 0806.1245}}].

\bibitem{brandenberger1989superstrings}
R.~Brandenberger and C.~Vafa, \emph{Superstrings in the early universe},
  {\emph{Nuclear Physics B} {\bfseries 316} (1989) 391}.

\bibitem{Brahma:2021tkh}
S.~Brahma, R.~Brandenberger and S.~Laliberte, \emph{{Emergent cosmology from
  matrix theory}}, \href{https://doi.org/10.1007/JHEP03(2022)067}{\emph{JHEP}
  {\bfseries 03} (2022) 067}
  [\href{https://arxiv.org/abs/2107.11512}{{\ttfamily 2107.11512}}].

\bibitem{Berera:1995ie}
A.~Berera, \emph{{Warm inflation}},
  \href{https://doi.org/10.1103/PhysRevLett.75.3218}{\emph{Phys. Rev. Lett.}
  {\bfseries 75} (1995) 3218}
  [\href{https://arxiv.org/abs/astro-ph/9509049}{{\ttfamily
  astro-ph/9509049}}].

\bibitem{Brahma:2024yor}
S.~Brahma, J.~Calder\'on-Figueroa and X.~Luo, \emph{{Time-convolutionless
  cosmological master equations: Late-time resummations and decoherence for
  non-local kernels}},  \href{https://arxiv.org/abs/2407.12091}{{\ttfamily
  2407.12091}}.

\bibitem{Brahma:2022yxu}
S.~Brahma, A.~Berera and J.~Calder\'on-Figueroa, \emph{{Quantum corrections to
  the primordial tensor spectrum: open EFTs \& Markovian decoupling of UV
  modes}}, \href{https://doi.org/10.1007/JHEP08(2022)225}{\emph{JHEP}
  {\bfseries 08} (2022) 225}
  [\href{https://arxiv.org/abs/2206.05797}{{\ttfamily 2206.05797}}].

\bibitem{Shandera:2017qkg}
S.~Shandera, N.~Agarwal and A.~Kamal, \emph{{Open quantum cosmological
  system}}, \href{https://doi.org/10.1103/PhysRevD.98.083535}{\emph{Phys. Rev.
  D} {\bfseries 98} (2018) 083535}
  [\href{https://arxiv.org/abs/1708.00493}{{\ttfamily 1708.00493}}].

\bibitem{Colas:2022kfu}
T.~Colas, J.~Grain and V.~Vennin, \emph{{Quantum recoherence in the early
  universe}},  \href{https://arxiv.org/abs/2212.09486}{{\ttfamily 2212.09486}}.

\bibitem{Burgess:2024eng}
C.P.~Burgess, T.~Colas, R.~Holman, G.~Kaplanek and V.~Vennin, \emph{{Cosmic
  purity lost: perturbative and resummed late-time inflationary decoherence}},
  \href{https://doi.org/10.1088/1475-7516/2024/08/042}{\emph{JCAP} {\bfseries
  08} (2024) 042} [\href{https://arxiv.org/abs/2403.12240}{{\ttfamily
  2403.12240}}].

\bibitem{Bhattacharyya_2024}
A.~Bhattacharyya, S.~Brahma, S.~Haque, J.S.~Lund and A.~Paul, \emph{The early
  universe as an open quantum system: complexity and decoherence},
  \href{https://doi.org/10.1007/jhep05(2024)058}{\emph{JHEP} {\bfseries 2024}
  (2024) }.

\bibitem{Bhattacharyya:2020rpy}
A.~Bhattacharyya, S.~Das, S.~Shajidul~Haque and B.~Underwood,
  \emph{{Cosmological Complexity}},
  \href{https://doi.org/10.1103/PhysRevD.101.106020}{\emph{Phys. Rev. D}
  {\bfseries 101} (2020) 106020}
  [\href{https://arxiv.org/abs/2001.08664}{{\ttfamily 2001.08664}}].

\bibitem{Bhattacharyya:2020kgu}
A.~Bhattacharyya, S.~Das, S.S.~Haque and B.~Underwood, \emph{{Rise of
  cosmological complexity: Saturation of growth and chaos}},
  \href{https://doi.org/10.1103/PhysRevResearch.2.033273}{\emph{Phys. Rev.
  Res.} {\bfseries 2} (2020) 033273}
  [\href{https://arxiv.org/abs/2005.10854}{{\ttfamily 2005.10854}}].

\bibitem{Haque:2021kdm}
S.S.~Haque, C.~Jana and B.~Underwood, \emph{{Saturation of thermal complexity
  of purification}}, \href{https://doi.org/10.1007/JHEP01(2022)159}{\emph{JHEP}
  {\bfseries 01} (2022) 159}
  [\href{https://arxiv.org/abs/2107.08969}{{\ttfamily 2107.08969}}].

\bibitem{Haque:2021hyw}
S.S.~Haque, C.~Jana and B.~Underwood, \emph{{Operator complexity for quantum
  scalar fields and cosmological perturbations}},
  \href{https://doi.org/10.1103/PhysRevD.106.063510}{\emph{Phys. Rev. D}
  {\bfseries 106} (2022) 063510}
  [\href{https://arxiv.org/abs/2110.08356}{{\ttfamily 2110.08356}}].

\bibitem{Lehners:2020pem}
J.-L.~Lehners and J.~Quintin, \emph{{Quantum Circuit Complexity of Primordial
  Perturbations}},
  \href{https://doi.org/10.1103/PhysRevD.103.063527}{\emph{Phys. Rev. D}
  {\bfseries 103} (2021) 063527}
  [\href{https://arxiv.org/abs/2012.04911}{{\ttfamily 2012.04911}}].

\bibitem{Liu:2021nzx}
L.-H.~Liu and A.-C.~Li, \emph{{Complexity of non-trivial sound speed in
  inflation}}, \href{https://doi.org/10.1016/j.dark.2022.101123}{\emph{Phys.
  Dark Univ.} {\bfseries 37} (2022) 101123}
  [\href{https://arxiv.org/abs/2102.12014}{{\ttfamily 2102.12014}}].

\bibitem{Saha:2022onq}
P.~Saha and M.~Park, \emph{{Primordial cosmic complexity and effects of
  reheating}}, \href{https://doi.org/10.1103/PhysRevD.108.083520}{\emph{Phys.
  Rev. D} {\bfseries 108} (2023) 083520}
  [\href{https://arxiv.org/abs/2212.13723}{{\ttfamily 2212.13723}}].

\bibitem{Li:2023ekd}
T.~Li and L.-H.~Liu, \emph{{Cosmological complexity of the modified dispersion
  relation}},  \href{https://arxiv.org/abs/2309.01595}{{\ttfamily 2309.01595}}.

\bibitem{NL1}
M.A.~Nielsen, \emph{{A geometric approach to quantum circuit lower bounds}},
  \href{https://doi.org/10.3390/universe5040092}{\emph{Science} {\bfseries 311}
  (2006) 92} [\href{https://arxiv.org/abs/0502070}{{\ttfamily 0502070}}].

\bibitem{NL2}
M.R.~Nielsen, M.~A.and~Dowling, M.~Gu and A.M.~Doherty, \emph{{Quantum
  Computation as Geometry}},
  \href{https://doi.org/10.3390/universe5040092}{\emph{Science} {\bfseries 311}
  (2006) 1133} [\href{https://arxiv.org/abs/0603161}{{\ttfamily 0603161}}].

\bibitem{NL3}
M.R.~Nielsen, M.~A.and~Dowling, \emph{{The geometry of quantum computation}},
  \href{https://doi.org/10.3390/universe5040092}{\emph{Science} {\bfseries 311}
  (2006) 1133} [\href{https://arxiv.org/abs/0701004}{{\ttfamily 0701004}}].

\bibitem{Jefferson}
R.~Jefferson and R.C.~Myers, \emph{{Circuit complexity in quantum field
  theory}}, \href{https://doi.org/10.1007/JHEP10(2017)107}{\emph{JHEP}
  {\bfseries 10} (2017) 107}
  [\href{https://arxiv.org/abs/1707.08570}{{\ttfamily 1707.08570}}].

\bibitem{Chapman:2017rqy}
S.~Chapman, M.P.~Heller, H.~Marrochio and F.~Pastawski, \emph{{Toward a
  Definition of Complexity for Quantum Field Theory States}},
  \href{https://doi.org/10.1103/PhysRevLett.120.121602}{\emph{Phys. Rev. Lett.}
  {\bfseries 120} (2018) 121602}
  [\href{https://arxiv.org/abs/1707.08582}{{\ttfamily 1707.08582}}].

\bibitem{Bhattacharyya:2018wym}
A.~Bhattacharyya, P.~Caputa, S.R.~Das, N.~Kundu, M.~Miyaji and T.~Takayanagi,
  \emph{{Path-Integral Complexity for Perturbed CFTs}},
  \href{https://doi.org/10.1007/JHEP07(2018)086}{\emph{JHEP} {\bfseries 07}
  (2018) 086} [\href{https://arxiv.org/abs/1804.01999}{{\ttfamily
  1804.01999}}].

\bibitem{Caputa:2017yrh}
P.~Caputa, N.~Kundu, M.~Miyaji, T.~Takayanagi and K.~Watanabe, \emph{{Liouville
  Action as Path-Integral Complexity: From Continuous Tensor Networks to
  AdS/CFT}}, \href{https://doi.org/10.1007/JHEP11(2017)097}{\emph{JHEP}
  {\bfseries 11} (2017) 097}
  [\href{https://arxiv.org/abs/1706.07056}{{\ttfamily 1706.07056}}].

\bibitem{Ali:2018fcz}
T.~Ali, A.~Bhattacharyya, S.~Shajidul~Haque, E.H.~Kim and N.~Moynihan,
  \emph{{Time Evolution of Complexity: A Critique of Three Methods}},
  \href{https://doi.org/10.1007/JHEP04(2019)087}{\emph{JHEP} {\bfseries 04}
  (2019) 087} [\href{https://arxiv.org/abs/1810.02734}{{\ttfamily
  1810.02734}}].

\bibitem{Bhattacharyya:2018bbv}
A.~Bhattacharyya, A.~Shekar and A.~Sinha, \emph{{Circuit complexity in
  interacting QFTs and RG flows}},
  \href{https://doi.org/10.1007/JHEP10(2018)140}{\emph{JHEP} {\bfseries 10}
  (2018) 140} [\href{https://arxiv.org/abs/1808.03105}{{\ttfamily
  1808.03105}}].

\bibitem{Hackl:2018ptj}
L.~Hackl and R.C.~Myers, \emph{{Circuit complexity for free fermions}},
  \href{https://doi.org/10.1007/JHEP07(2018)139}{\emph{JHEP} {\bfseries 07}
  (2018) 139} [\href{https://arxiv.org/abs/1803.10638}{{\ttfamily
  1803.10638}}].

\bibitem{Khan:2018rzm}
R.~Khan, C.~Krishnan and S.~Sharma, \emph{{Circuit Complexity in Fermionic
  Field Theory}}, \href{https://doi.org/10.1103/PhysRevD.98.126001}{\emph{Phys.
  Rev. D} {\bfseries 98} (2018) 126001}
  [\href{https://arxiv.org/abs/1801.07620}{{\ttfamily 1801.07620}}].

\bibitem{Camargo:2018eof}
H.A.~Camargo, P.~Caputa, D.~Das, M.P.~Heller and R.~Jefferson,
  \emph{{Complexity as a novel probe of quantum quenches: universal scalings
  and purifications}},
  \href{https://doi.org/10.1103/PhysRevLett.122.081601}{\emph{Phys. Rev. Lett.}
  {\bfseries 122} (2019) 081601}
  [\href{https://arxiv.org/abs/1807.07075}{{\ttfamily 1807.07075}}].

\bibitem{Ali:2018aon}
T.~Ali, A.~Bhattacharyya, S.~Shajidul~Haque, E.H.~Kim and N.~Moynihan,
  \emph{{Post-Quench Evolution of Complexity and Entanglement in a Topological
  System}}, \href{https://doi.org/10.1016/j.physletb.2020.135919}{\emph{Phys.
  Lett. B} {\bfseries 811} (2020) 135919}
  [\href{https://arxiv.org/abs/1811.05985}{{\ttfamily 1811.05985}}].

\bibitem{Caputa:2018kdj}
P.~Caputa and J.M.~Magan, \emph{{Quantum Computation as Gravity}},
  \href{https://doi.org/10.1103/PhysRevLett.122.231302}{\emph{Phys. Rev. Lett.}
  {\bfseries 122} (2019) 231302}
  [\href{https://arxiv.org/abs/1807.04422}{{\ttfamily 1807.04422}}].

\bibitem{Guo:2018kzl}
M.~Guo, J.~Hernandez, R.C.~Myers and S.-M.~Ruan, \emph{{Circuit Complexity for
  Coherent States}}, \href{https://doi.org/10.1007/JHEP10(2018)011}{\emph{JHEP}
  {\bfseries 10} (2018) 011}
  [\href{https://arxiv.org/abs/1807.07677}{{\ttfamily 1807.07677}}].

\bibitem{Bhattacharyya:2019kvj}
A.~Bhattacharyya, P.~Nandy and A.~Sinha, \emph{{Renormalized Circuit
  Complexity}},
  \href{https://doi.org/10.1103/PhysRevLett.124.101602}{\emph{Phys. Rev. Lett.}
  {\bfseries 124} (2020) 101602}
  [\href{https://arxiv.org/abs/1907.08223}{{\ttfamily 1907.08223}}].

\bibitem{Flory:2020eot}
M.~Flory and M.P.~Heller, \emph{{Geometry of Complexity in Conformal Field
  Theory}}, \href{https://doi.org/10.1103/PhysRevResearch.2.043438}{\emph{Phys.
  Rev. Res.} {\bfseries 2} (2020) 043438}
  [\href{https://arxiv.org/abs/2005.02415}{{\ttfamily 2005.02415}}].

\bibitem{Erdmenger:2020sup}
J.~Erdmenger, M.~Gerbershagen and A.-L.~Weigel, \emph{{Complexity measures from
  geometric actions on Virasoro and Kac-Moody orbits}},
  \href{https://doi.org/10.1007/JHEP11(2020)003}{\emph{JHEP} {\bfseries 11}
  (2020) 003} [\href{https://arxiv.org/abs/2004.03619}{{\ttfamily
  2004.03619}}].

\bibitem{Ali:2019zcj}
T.~Ali, A.~Bhattacharyya, S.S.~Haque, E.H.~Kim, N.~Moynihan and J.~Murugan,
  \emph{{Chaos and Complexity in Quantum Mechanics}},
  \href{https://doi.org/10.1103/PhysRevD.101.026021}{\emph{Phys. Rev. D}
  {\bfseries 101} (2020) 026021}
  [\href{https://arxiv.org/abs/1905.13534}{{\ttfamily 1905.13534}}].

\bibitem{Bhattacharyya:2019txx}
A.~Bhattacharyya, W.~Chemissany, S.~Shajidul~Haque and B.~Yan, \emph{{Towards
  the web of quantum chaos diagnostics}},
  \href{https://doi.org/10.1140/epjc/s10052-022-10035-3}{\emph{Eur. Phys. J. C}
  {\bfseries 82} (2022) 87} [\href{https://arxiv.org/abs/1909.01894}{{\ttfamily
  1909.01894}}].

\bibitem{Caceres:2019pgf}
E.~Caceres, S.~Chapman, J.D.~Couch, J.P.~Hernandez, R.C.~Myers and S.-M.~Ruan,
  \emph{{Complexity of Mixed States in QFT and Holography}},
  \href{https://doi.org/10.1007/JHEP03(2020)012}{\emph{JHEP} {\bfseries 03}
  (2020) 012} [\href{https://arxiv.org/abs/1909.10557}{{\ttfamily
  1909.10557}}].

\bibitem{Bhattacharyya:2020art}
A.~Bhattacharyya, W.~Chemissany, S.S.~Haque, J.~Murugan and B.~Yan, \emph{{The
  Multi-faceted Inverted Harmonic Oscillator: Chaos and Complexity}},
  \href{https://doi.org/10.21468/SciPostPhysCore.4.1.002}{\emph{SciPost Phys.
  Core} {\bfseries 4} (2021) 002}
  [\href{https://arxiv.org/abs/2007.01232}{{\ttfamily 2007.01232}}].

\bibitem{Liu_2020}
F.~Liu, S.~Whitsitt, J.B.~Curtis, R.~Lundgren, P.~Titum, Z.-C.~Yang et~al.,
  \emph{{Circuit complexity across a topological phase transition}},
  \href{https://doi.org/10.1103/PhysRevResearch.2.013323}{\emph{Phys. Rev.
  Res.} {\bfseries 2} (2020) 013323}
  [\href{https://arxiv.org/abs/1902.10720}{{\ttfamily 1902.10720}}].

\bibitem{Susskind:2020gnl}
L.~Susskind and Y.~Zhao, \emph{{Complexity and Momentum}},
  \href{https://arxiv.org/abs/2006.03019}{{\ttfamily 2006.03019}}.

\bibitem{Chen:2020nlj}
B.~Chen, B.~Czech and Z.-z.~Wang, \emph{{Cutoff Dependence and Complexity of
  the CFT$_2$ Ground State}},
  \href{https://arxiv.org/abs/2004.11377}{{\ttfamily 2004.11377}}.

\bibitem{Czech:2017ryf}
B.~Czech, \emph{{Einstein Equations from Varying Complexity}},
  \href{https://doi.org/10.1103/PhysRevLett.120.031601}{\emph{Phys. Rev. Lett.}
  {\bfseries 120} (2018) 031601}
  [\href{https://arxiv.org/abs/1706.00965}{{\ttfamily 1706.00965}}].

\bibitem{Chapman:2018hou}
S.~Chapman, J.~Eisert, L.~Hackl, M.P.~Heller, R.~Jefferson, H.~Marrochio
  et~al., \emph{{Complexity and entanglement for thermofield double states}},
  \href{https://doi.org/10.21468/SciPostPhys.6.3.034}{\emph{SciPost Phys.}
  {\bfseries 6} (2019) 034} [\href{https://arxiv.org/abs/1810.05151}{{\ttfamily
  1810.05151}}].

\bibitem{Geng:2019yxo}
H.~Geng, \emph{{$T\bar{T}$ Deformation and the Complexity=Volume Conjecture}},
  \href{https://doi.org/10.1002/prop.202000036}{\emph{Fortsch. Phys.}
  {\bfseries 68} (2020) 2000036}
  [\href{https://arxiv.org/abs/1910.08082}{{\ttfamily 1910.08082}}].

\bibitem{Guo:2020dsi}
M.~Guo, Z.-Y.~Fan, J.~Jiang, X.~Liu and B.~Chen, \emph{{Circuit complexity for
  generalized coherent states in thermal field dynamics}},
  \href{https://doi.org/10.1103/PhysRevD.101.126007}{\emph{Phys. Rev.}
  {\bfseries D101} (2020) 126007}
  [\href{https://arxiv.org/abs/2004.00344}{{\ttfamily 2004.00344}}].

\bibitem{Couch:2021wsm}
J.~Couch, Y.~Fan and S.~Shashi, \emph{{Circuit Complexity in Topological
  Quantum Field Theory}},  \href{https://arxiv.org/abs/2108.13427}{{\ttfamily
  2108.13427}}.

\bibitem{Erdmenger:2021wzc}
J.~Erdmenger, M.~Flory, M.~Gerbershagen, M.P.~Heller and A.-L.~Weigel,
  \emph{{Exact Gravity Duals for Simple Quantum Circuits}},
  \href{https://arxiv.org/abs/2112.12158}{{\ttfamily 2112.12158}}.

\bibitem{Chagnet:2021uvi}
N.~Chagnet, S.~Chapman, J.~de~Boer and C.~Zukowski, \emph{{Complexity for
  Conformal Field Theories in General Dimensions}},
  \href{https://doi.org/10.1103/PhysRevLett.128.051601}{\emph{Phys. Rev. Lett.}
  {\bfseries 128} (2022) 051601}
  [\href{https://arxiv.org/abs/2103.06920}{{\ttfamily 2103.06920}}].

\bibitem{Koch:2021tvp}
R.d.M.~Koch, M.~Kim and H.J.R.~Van~Zyl, \emph{{Complexity from spinning
  primaries}}, \href{https://doi.org/10.1007/JHEP12(2021)030}{\emph{JHEP}
  {\bfseries 12} (2021) 030}
  [\href{https://arxiv.org/abs/2108.10669}{{\ttfamily 2108.10669}}].

\bibitem{Banerjee:2022ime}
A.~Banerjee, A.~Bhattacharyya, P.~Drashni and S.~Pawar, \emph{{From CFTs to
  theories with Bondi-Metzner-Sachs symmetries: Complexity and
  out-of-time-ordered correlators}},
  \href{https://doi.org/10.1103/PhysRevD.106.126022}{\emph{Phys. Rev. D}
  {\bfseries 106} (2022) 126022}
  [\href{https://arxiv.org/abs/2205.15338}{{\ttfamily 2205.15338}}].

\bibitem{Bhattacharyya:2022ren}
A.~Bhattacharyya, G.~Katoch and S.R.~Roy, \emph{{Complexity of warped conformal
  field theory}},
  \href{https://doi.org/10.1140/epjc/s10052-023-11212-8}{\emph{Eur. Phys. J. C}
  {\bfseries 83} (2023) 33} [\href{https://arxiv.org/abs/2202.09350}{{\ttfamily
  2202.09350}}].

\bibitem{Bhattacharyya:2023sjr}
A.~Bhattacharyya and P.~Nandi, \emph{{Circuit complexity for Carrollian
  Conformal (BMS) field theories}},
  \href{https://doi.org/10.1007/JHEP07(2023)105}{\emph{JHEP} {\bfseries 07}
  (2023) 105} [\href{https://arxiv.org/abs/2301.12845}{{\ttfamily
  2301.12845}}].

\bibitem{Bhattacharyya:2022rhm}
A.~Bhattacharyya, T.~Hanif, S.S.~Haque and A.~Paul, \emph{{Decoherence,
  entanglement negativity, and circuit complexity for an open quantum system}},
  \href{https://doi.org/10.1103/PhysRevD.107.106007}{\emph{Phys. Rev. D}
  {\bfseries 107} (2023) 106007}
  [\href{https://arxiv.org/abs/2210.09268}{{\ttfamily 2210.09268}}].

\bibitem{Bhattacharyya:2021fii}
A.~Bhattacharyya, T.~Hanif, S.S.~Haque and M.K.~Rahman, \emph{{Complexity for
  an open quantum system}},
  \href{https://doi.org/10.1103/PhysRevD.105.046011}{\emph{Phys. Rev. D}
  {\bfseries 105} (2022) 046011}
  [\href{https://arxiv.org/abs/2112.03955}{{\ttfamily 2112.03955}}].

\bibitem{Bhattacharyya:2020iic}
A.~Bhattacharyya, S.S.~Haque and E.H.~Kim, \emph{{Complexity from the reduced
  density matrix: a new diagnostic for chaos}},
  \href{https://doi.org/10.1007/JHEP10(2021)028}{\emph{JHEP} {\bfseries 10}
  (2021) 028} [\href{https://arxiv.org/abs/2011.04705}{{\ttfamily
  2011.04705}}].

\bibitem{Craps:2023rur}
B.~Craps, M.~De~Clerck, O.~Evnin and P.~Hacker, \emph{{Integrability and
  complexity in quantum spin chains}},
  \href{https://arxiv.org/abs/2305.00037}{{\ttfamily 2305.00037}}.

\bibitem{Jaiswal:2021tnt}
N.~Jaiswal, M.~Gautam and T.~Sarkar, \emph{{Complexity, information geometry,
  and Loschmidt echo near quantum criticality}},
  \href{https://doi.org/10.1088/1742-5468/ac7aa6}{\emph{J. Stat. Mech.}
  {\bfseries 2207} (2022) 073105}
  [\href{https://arxiv.org/abs/2110.02099}{{\ttfamily 2110.02099}}].

\bibitem{Jaiswal:2020snm}
N.~Jaiswal, M.~Gautam and T.~Sarkar, \emph{{Complexity and information geometry
  in the transverse XY model}},
  \href{https://doi.org/10.1103/PhysRevE.104.024127}{\emph{Phys. Rev. E}
  {\bfseries 104} (2021) 024127}
  [\href{https://arxiv.org/abs/2005.03532}{{\ttfamily 2005.03532}}].

\bibitem{Bhattacharya:2022wlp}
A.~Bhattacharya, A.~Bhattacharyya and S.~Maulik, \emph{{Pseudocomplexity of
  purification for free scalar field theories}},
  \href{https://doi.org/10.1103/PhysRevD.106.086010}{\emph{Phys. Rev. D}
  {\bfseries 106} (2022) 086010}
  [\href{https://arxiv.org/abs/2209.00049}{{\ttfamily 2209.00049}}].

\bibitem{Chapman:2021jbh}
S.~Chapman and G.~Policastro, \emph{{Quantum Computational Complexity -- From
  Quantum Information to Black Holes and Back}},
  \href{https://arxiv.org/abs/2110.14672}{{\ttfamily 2110.14672}}.

\bibitem{Bhattacharyya:2021cwf}
A.~Bhattacharyya, \emph{{Circuit complexity and (some of) its applications}},
  \href{https://doi.org/10.1142/S0218301321300058}{\emph{Int. J. Mod. Phys. E}
  {\bfseries 30} (2021) 2130005}.

\bibitem{Katoch:2023etn}
G.~Katoch, \emph{{Investigations of LST and WCFT using complexity as a probe}},
  Ph.D. thesis, Indian Inst. Tech., Hyderabad, 8, 2023.

\bibitem{Aguilar-Gutierrez:2024yzu}
S.E.~Aguilar-Gutierrez, \emph{{De Sitter space, complexity, and the
  double-scaled SYK model}}, Ph.D. thesis, Leuven U., 2024.
\newblock \href{https://arxiv.org/abs/2406.19089}{{\ttfamily 2406.19089}}.

\bibitem{PhysRevLett.46.211}
A.O.~Caldeira and A.J.~Leggett, \emph{Influence of dissipation on quantum
  tunneling in macroscopic systems},
  \href{https://doi.org/10.1103/PhysRevLett.46.211}{\emph{Phys. Rev. Lett.}
  {\bfseries 46} (1981) 211}.

\bibitem{Martin:2021znx}
J.~Martin, A.~Micheli and V.~Vennin, \emph{{Discord and decoherence}},
  \href{https://doi.org/10.1088/1475-7516/2022/04/051}{\emph{JCAP} {\bfseries
  04} (2022) 051} [\href{https://arxiv.org/abs/2112.05037}{{\ttfamily
  2112.05037}}].

\bibitem{Martin:2022kph}
J.~Martin, A.~Micheli and V.~Vennin, \emph{{Comparing quantumness criteria}},
  \href{https://doi.org/10.1209/0295-5075/acc3be}{\emph{EPL} {\bfseries 142}
  (2023) 18001} [\href{https://arxiv.org/abs/2211.10114}{{\ttfamily
  2211.10114}}].

\bibitem{Assassi:2013gxa}
V.~Assassi, D.~Baumann, D.~Green and L.~McAllister, \emph{{Planck-Suppressed
  Operators}}, \href{https://doi.org/10.1088/1475-7516/2014/01/033}{\emph{JCAP}
  {\bfseries 01} (2014) 033} [\href{https://arxiv.org/abs/1304.5226}{{\ttfamily
  1304.5226}}].

\bibitem{Colas:2024xjy}
T.~Colas, C.~de~Rham and G.~Kaplanek, \emph{{Decoherence out of fire: purity
  loss in expanding and contracting universes}},
  \href{https://doi.org/10.1088/1475-7516/2024/05/025}{\emph{JCAP} {\bfseries
  05} (2024) 025} [\href{https://arxiv.org/abs/2401.02832}{{\ttfamily
  2401.02832}}].

\bibitem{Brahma:2024ycc}
S.~Brahma, J.~Calder\'on-Figueroa, X.~Luo and D.~Seery, \emph{{The special case
  of slow-roll attractors in de Sitter: Non-Markovian noise and evolution of
  entanglement entropy}},  \href{https://arxiv.org/abs/2411.08632}{{\ttfamily
  2411.08632}}.

\bibitem{Colas2022}
T.~Colas, J.~Grain and V.~Vennin, \emph{Benchmarking the cosmological master
  equations}, \href{https://doi.org/10.1140/epjc/s10052-022-11047-9}{\emph{The
  European Physical Journal C} {\bfseries 82} (2022) 1085}.

\bibitem{Burgess:2017ytm}
C.P.~Burgess, \emph{{Intro to Effective Field Theories and Inflation}},
  \href{https://arxiv.org/abs/1711.10592}{{\ttfamily 1711.10592}}.

\bibitem{Shiu:2011qw}
G.~Shiu and J.~Xu, \emph{{Effective Field Theory and Decoupling in Multi-field
  Inflation: An Illustrative Case Study}},
  \href{https://doi.org/10.1103/PhysRevD.84.103509}{\emph{Phys. Rev. D}
  {\bfseries 84} (2011) 103509}
  [\href{https://arxiv.org/abs/1108.0981}{{\ttfamily 1108.0981}}].

\bibitem{Pinol:2020kvw}
L.~Pinol, \emph{{Multifield inflation beyond $N_\mathrm{field}=2$:
  non-Gaussianities and single-field effective theory}},
  \href{https://doi.org/10.1088/1475-7516/2021/04/002}{\emph{JCAP} {\bfseries
  04} (2021) 002} [\href{https://arxiv.org/abs/2011.05930}{{\ttfamily
  2011.05930}}].

\bibitem{Pinol:2021aun}
L.~Pinol, S.~Aoki, S.~Renaux-Petel and M.~Yamaguchi, \emph{{Inflationary flavor
  oscillations and the cosmic spectroscopy}},
  \href{https://doi.org/10.1103/PhysRevD.107.L021301}{\emph{Phys. Rev. D}
  {\bfseries 107} (2023) L021301}
  [\href{https://arxiv.org/abs/2112.05710}{{\ttfamily 2112.05710}}].

\bibitem{Tolley:2007nq}
A.J.~Tolley and D.H.~Wesley, \emph{{Scale-invariance in expanding and
  contracting universes from two-field models}},
  \href{https://doi.org/10.1088/1475-7516/2007/05/006}{\emph{JCAP} {\bfseries
  05} (2007) 006} [\href{https://arxiv.org/abs/hep-th/0703101}{{\ttfamily
  hep-th/0703101}}].

\bibitem{Paris_2003}
M.G.A.~Paris, F.~Illuminati, A.~Serafini and S.~De~Siena, \emph{Purity of
  gaussian states: Measurement schemes and time evolution in noisy channels},
  \href{https://doi.org/10.1103/physreva.68.012314}{\emph{Physical Review A}
  {\bfseries 68} (2003) }.

\bibitem{Agon:2018zso}
C.A.~Ag\'on, M.~Headrick and B.~Swingle, \emph{{Subsystem Complexity and
  Holography}}, \href{https://doi.org/10.1007/JHEP02(2019)145}{\emph{JHEP}
  {\bfseries 02} (2019) 145}
  [\href{https://arxiv.org/abs/1804.01561}{{\ttfamily 1804.01561}}].

\bibitem{Ghodrati:2019hnn}
M.~Ghodrati, X.-M.~Kuang, B.~Wang, C.-Y.~Zhang and Y.-T.~Zhou, \emph{{The
  connection between holographic entanglement and complexity of purification}},
  \href{https://doi.org/10.1007/JHEP09(2019)009}{\emph{JHEP} {\bfseries 09}
  (2019) 009} [\href{https://arxiv.org/abs/1902.02475}{{\ttfamily
  1902.02475}}].

\bibitem{Bhattacharyya:2018sbw}
A.~Bhattacharyya, T.~Takayanagi and K.~Umemoto, \emph{{Entanglement of
  Purification in Free Scalar Field Theories}},
  \href{https://doi.org/10.1007/JHEP04(2018)132}{\emph{JHEP} {\bfseries 04}
  (2018) 132} [\href{https://arxiv.org/abs/1802.09545}{{\ttfamily
  1802.09545}}].

\bibitem{Colas:2021llj}
T.~Colas, J.~Grain and V.~Vennin, \emph{{Four-mode squeezed states: two-field
  quantum systems and the symplectic group $\mathrm {Sp}(4,{\mathbb {R}})$}},
  \href{https://doi.org/10.1140/epjc/s10052-021-09922-y}{\emph{Eur. Phys. J. C}
  {\bfseries 82} (2022) 6} [\href{https://arxiv.org/abs/2104.14942}{{\ttfamily
  2104.14942}}].

\end{thebibliography}\endgroup
\end{document}